\documentclass[aps,prl,twocolumn,nofootinbib,superscriptaddress]{revtex4-1}
\usepackage[a4paper,top=2.00cm,bottom=2.00cm,left=2.00cm,right=2.00cm]{geometry}
\usepackage{amsfonts}
\usepackage{amscd}
\usepackage{amsmath}
\usepackage{amssymb}
\usepackage{txfonts}
\usepackage{graphicx}
\usepackage{dcolumn}
\usepackage{bm}
\usepackage{xcolor}
\usepackage[%
  colorlinks=true,
  urlcolor=blue,
  linkcolor=blue,
  citecolor=blue
]{hyperref}
\usepackage{tabularx}
\usepackage{etoolbox}
\newcommand{\nn}{\nonumber}
\newcommand{\beq}{\begin{eqnarray}}
\newcommand{\eeq}{\end{eqnarray}}

\def\gauge{{\tt U(1)_{\text{N}}}} 
\def\so2{{\tt SO(2)}}
 
\def\time{{\tt T}}
\def\d3d{{\tt D_{3d}}}

\def\c2v{{\tt C_{2v}}}
\makeatletter
\let\cat@comma@active\@empty
\makeatother
\begin{document}
\title{Chiral Higgs Mode in Nematic Superconductors}
\author{Hiroki Uematsu}
\affiliation{Department of Materials Engineering Science, Osaka University, Toyonaka, Osaka 560-8531, Japan}
\author{Takeshi Mizushima}
\email{mizushima@mp.es.osaka-u.ac.jp}
\affiliation{Department of Materials Engineering Science, Osaka University, Toyonaka, Osaka 560-8531, Japan}
\author{Atsushi Tsuruta}
\affiliation{Department of Materials Engineering Science, Osaka University, Toyonaka, Osaka 560-8531, Japan}
\author{Satoshi Fujimoto}
\affiliation{Department of Materials Engineering Science, Osaka University, Toyonaka, Osaka 560-8531, Japan}
\author{J. A. Sauls}
\affiliation{
Center for Applied Physics \& Superconducting Technologies
\\
Department of Physics, Northwestern University, Evanston, IL 60208 USA
}

\date{\today}

\begin{abstract}
Nematic superconductivity with spontaneously broken rotation symmetry has recently been reported in doped topological insulators, $M_x$Bi$_2$Se$_3$ ($M$=Cu, Sr, Nb). Here we show that the electromagnetic (EM) response of these compounds provides a spectroscopy for bosonic excitations that reflect the pairing channel and the broken symmetries of the ground state. Using quasiclassical Keldysh theory, we find two characteristic bosonic modes in nematic superconductors: the nematicity mode and the chiral Higgs mode. The former corresponds to the vibrations of the nematic order parameter associated with broken crystal symmetry, while the latter represents the excitation of chiral Cooper pairs. The chiral Higgs mode softens at a critical doping, signaling a dynamical instability of the nematic state towards a new chiral ground state with broken time reversal and mirror symmetry. Evolution of the bosonic spectrum is directly captured by EM power absorption spectra. We also discuss contributions to the bosonic spectrum from sub-dominant pairing channels to the EM response.
\end{abstract}

\maketitle

\begin{figure}[t!]
\includegraphics[width=85mm]{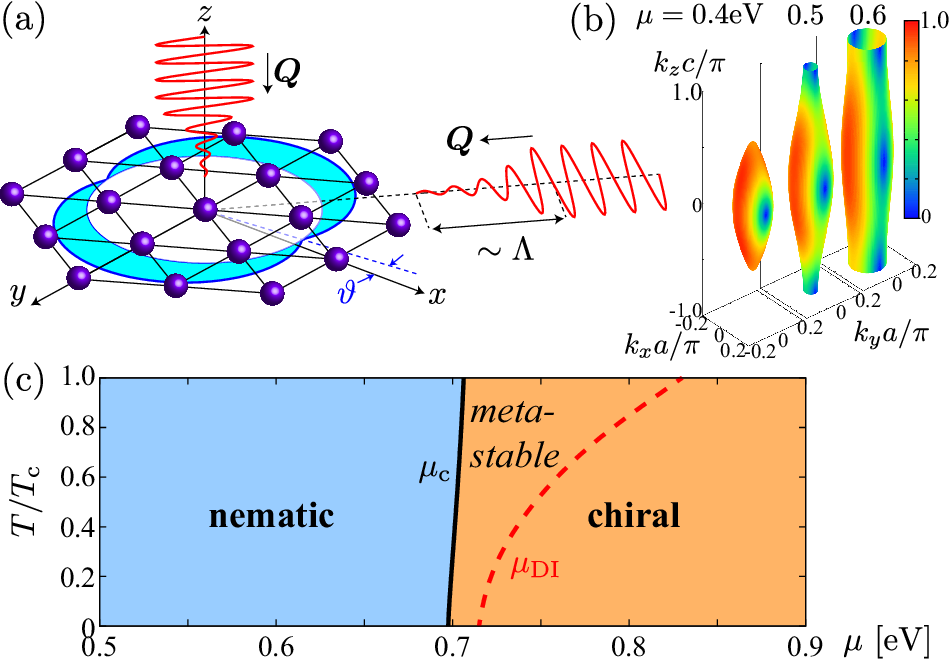}
\caption{(a) Configurations of polarized EM waves to probe the nematic pairing gap in the $\d3d$ crystal structure. (b) Evolution of the Fermi surface and superconducting gap in the nematic state for various $\mu$. (c) Phase diagram of $M_x$Bi$_2$Se$_3$ computed by the quasiclassical theory with $\eta _1 = \Delta(T,\mu)$ and $\eta _2=0$. We set $T^{(E_u,2)}_{\rm c}=T_{\rm c}$ with $T_{\rm c} \equiv T^{(E_u,1)}_{\rm c}$. The dashed curve shows the dynamical instability of the chiral Higgs mode beyond which the nematic state is no longer metastable.}
\label{fig:lattice}
\end{figure}

{\it Introduction.} Spontaneous symmetry breaking is an important concept that spreads across the diverse fields of modern physics. The recent discovery of two-fold rotation symmetry in superconducting compounds, $M_x$Bi$_2$Se$_3$ ($M={\rm Cu}$, Sr, Nb), has stimulated an intense discussion of superconductivity with a new class of spontaneous symmetry breaking~\cite{matano,yonezawa,panSR16,nikitinPRB16,asabaPRX17,shenNPJ17,du17,smylie17,smylie18,kuntsevicha18,willa}. The rotation symmetry breaking in the basal plane is compatible with odd-parity time-reversal invariant pairing belonging to the two-dimensional irreducible representation ($E_u$) of the $\d3d$ symmetry, which exhibits twofold symmetric gap anisotropy (Fig.~\ref{fig:lattice}). The anisotropy is represented by a nematic order parameter~\cite{fuPRB14}. The odd-parity superconductor (SC) $M_x$Bi$_2$Se$_3$ has also attracted much attention as a prototype of DIII topological SCs that host helical Majorana fermions~\cite{satoPRB10,fuPRL10,sasakiPRL11,yamakage12,hao11,hsieh12,yipPRB13,mizushimaPRB14,sasakiPC15}. In addition, there exist competing pairing channels corresponding to the $A_{1g}$, $A_{1u}$, and $A_{2u}$ irreducible representations, in addition to the ``nematic'' $E_u$ state~\cite{fuPRL10}.

In this Letter, we report theoretical results showing that the electromagnetic (EM) response at microwave frequencies provides a spectroscopy for long-lived bosonic excitations that are ``fingerprints'' of the nematic ground state that breaks the maximal symmetry ${\tt G}=\d3d\times \time \times \gauge$ of the parent compound down to ${\tt H=\c2v\times\time}$, where $\time$, $\gauge$, and $\d3d$ and $\c2v$ denote time-reversal symmetry, global gauge symmetry, and point-groups for three- and two-fold rotations, respectively~\cite{landau}. We first discuss the Fermi-surface evolution that drives the nematic-to-chiral phase transition within the $E_u$ representation. Using the quasiclassical Keldysh theory, we find two characteristic bosonic modes in nematic SCs: the nematicity mode and the chiral Higgs mode. The former corresponding to transverse oscillations of the nematic order parameter is the pseudo-Nambu-Goldstone (NG) boson associated with the broken $\d3d$ symmetry. The latter represents the excitation of chiral Cooper pairs. We find that the mass gap of the chiral mode tends to zero as the Fermi surface changes topology from a closed spherical shape to an open cylindrical Fermi surface, signaling the dynamical instability of the nematic state towards the chiral state with broken time-reversal and mirror symmetries. Bosonic modes of unconventional SCs involve the coherent dynamics of macroscopic fractions of electrons, and reflect the broken symmetries and the sub-dominant pairing interactions~\cite{hirschfeld,hirschfeldPRL92,yip,sauls15,hao18,haoPhD,royPRB08,higashitaniPRB00,miura,hirashima,monienJLTP86,fayPRB00,hirschfeldPRB15,bittnerPRL15,lutchynPRB08}. The bosonic excitation spectrum can be detected through transverse EM wave absorption [see Fig.~\ref{fig:lattice}(a)]. We also consider bosonic modes corresponding to the sub-dominant odd-parity $A_{1u}$ and $A_{2u}$ representations.

{\it Effective Hamiltonian.} Electrons embedded in $M_x$Bi$_2$Se$_3$ exhibit (i) the orbital degrees of freedom, (ii) strong spin-orbit coupling, and (iii) evolution of the Fermi surface with increase in carrier concentration (see Fig.~\ref{fig:lattice})~\cite{lahoudPRB13,lawsonPRB14}. The low-energy physics is governed by electrons in two $p_z$-orbitals near the Fermi level. The effective Hamiltonian is given as~\cite{zhangNP09,LiuPRB10,hashimotoJPSJ13,haoPRB14}
\begin{align}
\xi ({\bm p}) =& c({\bm p}) + m({\bm p})\sigma _x + v_z f_z(p_z)\sigma _y 
+ v\left( {\bm p}\times {\bm s} \right)_z\sigma _z -\mu,
\label{eq:htik}
\end{align}
where $c ({\bm p})=c_0+c_1 f_{\perp}({\bm p})+c_2p^2_{\parallel}$, $m ({\bm p})=m_0+m_1 f_{\perp}({\bm p})+m_2p^2_{\parallel}$, and $\mu$ are the diagonal self-energy correction, band gap, and the chemical potential, respectively ($p^2_{\parallel}\equiv p^2_x+p^2_y$). Nearest-neighbor hopping along the $z$ direction gives $f_z(p_z)=\frac{1}{c} \sin(p_zc)$ and $f_{\perp} =\frac{2}{c^2}[1-\cos(p_zc)]$. We take the $\hat{\bm z}$ axis along the $(111)$ direction of the crystal, and ${\bm s}$ (${\bm \sigma}$) is the spin (orbital) Pauli matrices. The Hamiltonian in Eq.~\eqref{eq:htik} maintains the enlarged ${\tt D_{\infty}}$ symmetry including $\so2$ about the $z$-axis, while a higher order correction on $p$ introduces three mirror planes and threefold rotational symmetry in the $xy$ plane~\cite{fuPRB14}.

The intercalation of $M$ atoms increases the carrier concentration in the conduction band (CB). As $\mu \gg \Delta$ in typical materials (where $\Delta$ is the superconducting gap), low-energy properties of the superconducting states are governed by the CB electrons with the disperion $E_{\rm CB}({\bm p})=c-\mu+\sqrt{m^2+v^2_zf^2_{z}+v^2p^2_{\parallel}}$, which is well separated from the valence band by the band gap, $|m_0|\sim \mu$, at the $\Gamma$ point. Hence, we focus on the Hamiltonian for CB electrons interacting through the odd-parity pairing interaction within $\d3d\times\time\times\gauge$,

\begin{align}
V_{\mu \nu}({\bm p},{\bm p}^{\prime})
= -\sum^{\rm odd}_{\Gamma}\sum^{n_{\Gamma}}_{i=1}
V^{(\Gamma)}_i{d}^{\Gamma}_{\mu,i}({\bm p}){d}^{\Gamma\ast}_{\nu,i}({\bm p}^{\prime}).
\label{eq:int}
\end{align}
where $\Gamma = A_{1u}$, $A_{2u}$, and $E_u$ are the odd-parity irreducible representations of $\d3d$ with the dimension $n_{\Gamma}$ and basis functions $\{{\bm d}^{{\Gamma}}_{1},\cdots,{\bm d}^{{\Gamma}}_{n_{\Gamma}}\}$. The basis functions in lowest order in $p$ are ${\bm d}^{E_u}_1=(v_zf_z/|m_0|,0,-vp_x/m_0)$, ${\bm d}^{E_u}_2=(0,v_zf_z/|m_0|,-vp_y/m_0)$, ${\bm d}^{A_{1u}}_1=(vp_x/|m_0|, vp_y/|m_0|, vf_z({\bm p})/m_0)$, and ${\bm d}^{A_{2u}}_1=( - vp_y, vp_x, 0)/|m_0|$. In the following we utilize the more general form of ${\bm d}^{\Gamma}_i$~\cite{SM}. By employing the regularization of gap equations, the pairing interaction of the $(\Gamma,i)$ channel, $V^{(\Gamma)}_i$, can be related to the instability temperature of the $(\Gamma,i)$ gap function, $T^{(\Gamma,i)}_{\rm c}$~\cite{SM}. We set $T_{\rm c}\equiv T^{(E_u,1)}_{\rm c}$.

{\it Nematic-to-chiral phase transition.} We consider the ground state within the $E_u$ representation, i.e., $T_{\rm c}\!\ge \! T^{(E_u,2)}_{\rm c} \!>\! T^{(A_{1u})}_{\rm c}, T^{(A_{2u})}_{\rm c}$, where the equilibrium odd-parity $E_u$ order parameter in the CB is given by 
\begin{equation}
{\bm d}({\bm p})=\eta _1 {\bm d}^{E_u}_1({\bm p}) + \eta _2  {\bm d}^{E_u}_2({\bm p})
\,.
\label{eq:eta}
\end{equation}
The nematic state with $(\eta _1,\eta_2) = \Delta (\cos\vartheta,\sin\vartheta)$ spontaneously breaks rotational symmetry, and is degenerate with respect to the angle $\vartheta \in [0,\pi/2]$. The broken symmetry is characterized by nematic order, $Q\equiv (|\eta _1|^2-|\eta _2|^2,\eta _1 \eta^{\ast}_2 + \eta^{\ast}_1\eta _2)$~\cite{fuPRB14,venderbosPRB16}. The angle {$\vartheta$} represents the orientation of two point nodes in the $xy$ plane (Fig.~\ref{fig:lattice}). Although Eq.~\eqref{eq:htik} respects ${\tt D_{\infty}}$ symmetry, corrections to Eq.~\eqref{eq:htik} from hexagonal warping of the Fermi surface pins the nematic angle $\vartheta$ to one of three equivalent crystal axes. Another competing order {allowed} by Eq.~\eqref{eq:eta} is the chiral state with broken time-reversal symmetry, $(\eta_1, \eta_2)=\Delta(1,\pm i)$. The chiral state, ${\bm d}_1 \pm i{\bm d}_2$, is a non-unitary state with two distinct gaps: one full gap, and another with point nodes at ${\bm p}=\pm{p}_{{\rm F},z}\hat{\bm z}$.

In Fig.~\ref{fig:lattice}(c), we show the phase diagram of $M_x$Bi$_2$Se$_3$ obtained from quasiclassical theory~\cite{SM}. The intercalation of $M$ atoms between the quintuple layers modifies the $c$-axis length of the crystal, namely, the hopping parameters along the $z$-axis $(c_1, m_1, v_z)$. This makes the Fermi pocket around the $\Gamma$ point elongate in the $\hat{\bm z}$ direction. The Fermi surface indeed evolves from a closed spherical shape to a quasi-two-dimensional open cylinder as $\mu$ increases~\cite{lahoudPRB13,lawsonPRB14}. The gap structure of the nematic state changes from a point-nodal to a line-nodal structure as the Fermi surface evolves [ Fig.~\ref{fig:lattice}(b)]~\cite{hashimoto14}. In contrast, the point nodes of the chiral state disappear and the fully gapped chiral state becomes thermodynamically stable when the Fermi surface is opened in the $z$-direction. To incorporate the Fermi surface evolution, we follow Ref.~\cite{hashimoto14}: the set of parameters in Ref.~\cite{hashimotoJPSJ13} for $\mu = 0.4$eV and the half-value of $(c_1, m_1, v_z)$ for $\mu = 0.65$eV. The parameters for arbitrary $\mu$ are given by interpolating $(c_1, m_1, v_z)$ linearly with respect to $\mu$. With this parametrization, the Fermi surface is opened along the $z$-axis for $\mu \gtrsim 0.5$eV. Using this set of parameters, we calculate the thermodynamic potential within the quasiclassical theory, which is valid for $\Delta\ll\mu $. Figure~\ref{fig:lattice} shows the first-order phase boundary between the nematic and chiral ground states near $\mu _{\rm c}\sim0.7$eV. Thus, the nematic-to-chiral phase transition  can be driven by Fermi surface evolution, as well as the exchange coupling to magnetic moments of dopant atoms~\cite{yuanPRB17,chirolliPRB17} and the thickness of materials~\cite{zyuzinPRL17,chirolliPRB18}. Note that the result obtained above is based on a simple interpolation of the Fermi surface evolution. The phase boundary may be shifted in real materials.

{\it Nematicity and chiral Higgs mode.} Consider the nematic state, ${\bm d}({\bm p}_{\rm F}) = \Delta (T,\mu){\bm d}^{E_u}_1({\bm p}_{\rm F})$, corresponding to $\vartheta=0$. The fluctuations in the $E_u$ ground state, $\delta {\bm d}({\bm p}_{\rm F},{\bm Q},t) \equiv {\bm d}({\bm p}_{\rm F},{\bm Q},t)-\Delta {\bm d}^{E_u}_1({\bm p}_{\rm F})$, decompose into the $(\Gamma,j)$ eigenmodes 
\begin{equation}
\delta d^{\rm C}_{\mu}({\bm p}_{\rm F},{\bm Q},t)=\sum^{\rm odd}_{\Gamma} \sum^{n_{\Gamma}}_{j=1} \mathcal{D}^{\rm C}_{\Gamma,j}({\bm Q},t)d^{\Gamma}_{\mu,j} ({\bm p}_{\rm F}),
\label{eq:deltad}
\end{equation}
where ${\bm Q}$ is the center-of-mass momentum of Cooper pairs. In the weak-coupling limit, all the bosonic excitations are classified in terms of the parity under particle-hole conversion (${\rm C}=\pm$), $\mathcal{D}^{\rm C}_j=\mathcal{D}_j+{\rm C}\mathcal{D}^{\ast}_j$. The $A_{1u}$ and $A_{2u}$ modes may also exist as long-lived bosons in the spectrum of the nematic state even when $T_{\rm c}>T^{A_{1u}}_{\rm c}, T^{A_{2u}}_{\rm c}$. For $\Gamma=E_u$, there exist four collective modes. Two of these modes are fluctuations in the ground state sector, $\mathcal{D}^{\rm C}_{E_u,1}$, the other two modes are in the orthogonal sector $\mathcal{D}^{\rm C}_{E_u,2}$. 
The $\mathcal{D}^{\rm C}_{E_u,1}$ modes correspond to the NG mode associated with the broken $\gauge$ symmetry ($\mathcal{D}^-_{E_u,1}$), which is gapped out by the Anderson-Higgs mechanism~\cite{anderson63,higgs64}, and $\mathcal{D}^{\rm +}_{E_u,1}$ corresponding to the amplitude Higgs mode with mass $2\Delta$.

The bosonic modes orthogonal to the ground-state sector are represented by $\mathcal{D}^{\rm C}_{E_u,2}$. Let us define $\mathcal{D}^+_{E_u,2}=\mathcal{D}_{E_u,2}+\mathcal{D}^{\ast}_{E_u,2} = \Delta \delta \vartheta(t)$ and $\mathcal{D}^-_{E_u,2}=\mathcal{D}_{E_u,2}-\mathcal{D}^{\ast}_{E_u,2}=i\epsilon(t) \Delta$, where $\delta\vartheta, \epsilon \in \mathbb{R}$. Thus, $\mathcal{D}^+_{E_u,2}$ corresponds to ${\bm d}(t)=\Delta[{\bm d}^{E_u}_1+\delta\vartheta (t){\bm d}^{E_u}_2]$, for $|\delta \theta|\ll 1$. This is the pseudo-NG mode associated with the broken rotational symmetry, and represents fluctuations of the nematic order parameter $Q$. The $\mathcal{D}^-_{E_u,2}$ mode represents excitation of chiral Cooper pairs, ${\bm d}(t)= \Delta[{\bm d}^{E_u}_1+i\epsilon(t) {\bm d}^{E_u}_2]$. We refer to $\mathcal{D}^+_{E_u,2}$ and $\mathcal{D}^-_{E_u,2}$ as the nematicity mode and chiral Higgs mode, respectively. 

Let us now consider the linear response to EM fields, $-\frac{e}{c}{\bm v}_{\rm F}\cdot{\bm A}({\bm Q},\omega)$, where ${\bm A}$ is a vector potential. The dynamical properties of the superconducting state of $M_x$Bi$_2$Se$_3$ are governed by Bogoliubov quasiparticles (QPs) in the CB and long-lived bosonic excitations of the pair condensate. The bosonic excitations involve a coherent motion of macroscopic fractions of particles, while low-lying QPs are responsible for the dissipation and the pair-breaking channels. To incorporate the interplay between them, we utilize the quasiclassical Keldysh transport theory~\cite{serene}. The fundamental quantity is the quasiclassical Keldysh propagator for CB electrons, 
{which contains both Bogoliubov QPs and dynamical bosonic fields, and} are governed by the transport-like equation~\cite{eil68,serene,saulsPRB17}. 
The linear response of the order parameter to the vector potential ${\bm A}$ is obtained from 
the equations of motion 
\begin{align}
\left[
\omega^2 -\mathbb{M}^{\rm C}_{\Gamma,j}({\bm Q},\omega)
\right]
\mathcal{D}^{\rm C}_{\Gamma,j}({\bm Q},\omega)
= \frac{e}{c} Q_{\mu} \zeta^{(\Gamma,j)}_{\mu\nu}({\bm Q},\omega) A_{\nu} ,
\label{eq:eom}
\end{align}
where $\mathbb{M}^{-}_{\Gamma,j}$ microscopically determines the mass and lifetime of the mode~\cite{SM}. Note that particle-hole symmetry prohibits the direct coupling of the ${\rm C}=+$ nematicity mode to transverse EM fields (i.e., $\zeta =0$). However, the ${\rm C}=+$ mode does contribute to the dynamical spin susceptibility~\cite{SM}. For ${\rm C}=-$, the coupling of the EM field to the bosonic excitations is governed by the matrix elements
\begin{equation}
\zeta^{(\Gamma,j)}_{\mu\nu}({\bm Q},\omega)=\Delta \frac{\langle\bar{\lambda}({\bm p}_{\rm F},Q)v^{\mu}_{\rm F}v^{\nu}_{\rm F}{\bm d}^{(E_u)}_{1}({\bm p}_{\rm F})\cdot {\bm d}^{(\Gamma)}_{j}({\bm p}_{\rm F})\rangle_{\rm FS}}{\langle \bar{\lambda}({\bm p}_{\rm F},Q)|{\bm d}^{(\Gamma)}_{j} ({\bm p}_{\rm F})|\rangle_{\rm FS}}
\,,
\label{eq:zeta}
\end{equation}
where $\langle ... \rangle _{\rm FS} \equiv\int dS_{\bm p} ... $ is an average over the Fermi surface obtained from Eq.~\eqref{eq:htik} that satisfies $\int dS_{\bm p} = 1$. The {tensor} $v^{\mu}_{\rm F}v^{\nu}_{\rm F}$ {determines the coupling to} ${\bm A}$ ($v^{\nu}_{\rm F}$) and ${\bm Q}$ ($v^{\mu}_{\rm F}$), respectively, where $v^{\mu}_{\rm F}\equiv \partial E_{\rm CM}/\partial p_{\mu}$ is the Fermi velocity of CB electrons. Hence, $\zeta _{\mu\nu}^{(\Gamma,j)}$ in Eq.~\eqref{eq:zeta} determines the coupling of $(\Gamma,j)$ bosonic modes to EM fields with ${\bm A}$ and ${\bm Q}$. The generalized Tsuneto function~\cite{tsuneto}, given by $\bar\lambda= \int^{\infty}_{|{\bm d}|}\frac{d\epsilon}{\sqrt{\epsilon^2-|{\bm d}|^2}}\frac{\tanh(\epsilon/2T)}{\epsilon^2-\omega^2/4}$ {for} ${\bm Q}\rightarrow {\bm 0}$, is real and positive below the pair-breaking edge $\omega < |{\bm d}({\bm p}_{\rm F})|$, while it has an imaginary part for $\omega > |{\bm d}({\bm p}_{\rm F})|$ which contributes to the dissociation of bosonic modes into Bogoliubov QPs~\cite{SM,mck90}.

\begin{figure}[t!]
\includegraphics[width=85mm]{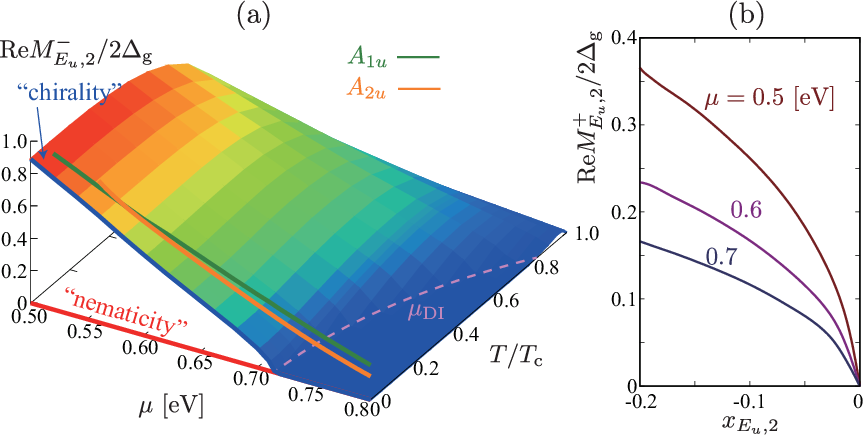}
\caption{
(a) Mass gap of the chirality mode ($M^-_{E_u,2}$), the nematicity mode ($M^+_{E_u,2}(0,\mu)$), and the $A_{1u}$ and $A_{2u}$ modes as a function of $\mu$. We set $x_{A_{1u}}=x_{A_{2u}}=-1.5$ and $x_{E_u,2}=0$, where $x_{\Gamma,j}=\ln T^{(\Gamma,j)}_{\rm c}/T_{\rm c}$. The color map shows the $T$-dependence of the chirality mode. The dashed curve corresponds to the dynamical instability of the chirality mode at which the mass gap closes. (b) Mass gap of the nematicity mode as a function of {extrinsic symmetry breaking of $E_u$ representation measured by} $x_{E_u,2}=\ln(T^{(E_u,2)}_{\rm c}/T_{\rm c})$.
}
\label{fig:mass}
\end{figure}

The dispersion relations, $\omega^{\rm C}_{\Gamma,j}({\bm Q})$, {are} determined from Eq.~\eqref{eq:eom} by solving the nonlinear equation, $\omega^2-\mathbb{M}^{\rm C}_{\Gamma,j}{({\bm 0},\omega)}=0$, which corresponds to a pole of $\delta \mathcal{D}^{\rm C}_{\Gamma,j}/\delta A_{\mu}$. In Fig.~\ref{fig:mass}, we plot the mass gap of {the} bosonic modes, $M^{\rm C}_{\Gamma,j}\equiv\omega^{\rm C}_{\Gamma,j}({\bm 0})$, including the chiral Higgs and nematicity modes. The parameters are the same as those in Fig.~\ref{fig:lattice}(c). At $T=0$, the nematicity mode remains gapless irrespective of $\mu$. The gapless spectrum of the nematicity mode is protected by the enlarged ${\tt D_{\infty}}$ symmetry of Eq.~\eqref{eq:htik}, and it is gapped out by terms that are higher-order {in} $p$, such as the hexagonal warping {energy}. Figure~\ref{fig:mass}(b) shows that the mass of the nematicity mode is sensitive to the splitting of $T_{\rm c}$ of the nematic $E_u$ states.

In Fig.~\ref{fig:mass}, the mass of the chiral Higgs mode decreases as $\mu$ increases and softens at the critical value $\mu_{\rm DI}=0.71$ at $T=0$. The softening indicates the dynamical instability of the nematic state towards the chiral state. As shown in Fig.~\ref{fig:lattice}(c), the dynamical instability at $T=0$ takes place in the vicinity of the nematic-to-chiral phase transition $\mu_{\rm c}$, while it deviates from $\mu_{\rm c}(T)$ with increasing $T$. This implies that $\mu _{\rm c}$ is the weak first-order transition in low temperatures and the softening can be indeed captured in experiments. The damping of the chirality mode is $-{\rm Im}M^-_{Eu,2}/2\Delta_{\rm g}=0.08$ at $\mu =0.5$eV. The chirality mode has a long lifetime for large $\mu$ [see Fig.~S3(a) in Ref.~\cite{SM}]. The QP density of states due to the point nodes decreases as $\omega^2$, which suppresses the pair-breaking channels for the decay of the chirality mode into QPs residing around the nodal points. Figure~\ref{fig:mass} also shows that the masses of bosonic modes supported by the competing pairing channels ($A_{1u}$ and $A_{2u}$) soften and their fluctuations develop as $\mu$ increases. 

{\it Selection rules and EM absorption spectra.} The signatures of the bosonic spectrum and its evolution, inherent to nematic SCs, is reflected in the microwave power absorption spectrum, $P(\omega)=\int d{\bm Q}{\rm Re}[{\bm j}({\bm Q},\omega)\cdot{\bm E}^{\ast}({\bm Q},\omega)]$, that is, the Joule losses of the electric field (${\bm E}$) and current (${\bm j}$) within the penetration depth $\Lambda=\sqrt{mc^2/4\pi ne^2}$~\cite{hirschfeld,yip,sauls15,hao18,haoPhD}. The charge current density is obtained from the quasiclassical propagator as~\cite{SM}
%
%
\begin{align}
\delta j_{\mu}(Q)
=& \sum _{\nu=x,y,z}\left\{ K^{\rm QP}_{\mu\nu} 
- eN_{\rm F}Q _{\tau}
\sum^{\rm odd}_{\Gamma}\sum^{n_{\Gamma}}_{j=1}\zeta^{(\Gamma)}_{\mu\tau,j}
\left(\frac{\delta\mathcal{D}^{-}_{\Gamma,j}}{\delta A_{\nu}}\right)
\right\}A_{\nu}.
\label{eq:currentK}
\end{align}
Equation~\eqref{eq:currentK} is the paramagnetic response function, including the vertex corrections from polarization of the medium by bosonic fields~\cite{hirschfeld}. The term, $K^{\rm QP}_{\mu\nu}=-\frac{2e^2N_{\rm F}}{c} \langle \{1+\frac{({\bm v}_{\rm F}\cdot{\bm Q})^2(1-\lambda)}{\omega^2-({\bm v}_{\rm F}\cdot{\bm Q})^2}\}{v}^{\mu}_{\rm F}{v}^{\nu}_{\rm F}\rangle_{\rm FS}$, where $\lambda = |{\bm d}|^2\bar{\lambda}$, describes the QP contribution to the dissipation via pair-breaking processes. 
The response, $\delta\mathcal{D}^{-}_{\Gamma,j}/\delta A_{\nu}$, is obtained from Eq.~\eqref{eq:eom}, which has a pole at the collective mode frequency $\omega^{-}_{\Gamma,j}({\bm Q})$ that satisfies $\omega^2-\mathbb{M}^{-}_{\Gamma,j}(\omega)=0$. 
For $Q v_{\rm F}\ll \Delta$ and $v_{\rm F}/\Lambda \ll \Delta$, the power absorption spectrum is decomposed into the QP contribution and a resonance part from the collective excitations, $P(\omega)=P^{\rm QP}(\omega)+P^{\rm CM}(\omega)$~\cite{SM}.

Equation \eqref{eq:zeta} determines the coupling of bosonic modes of the nematic state with ${\bm d}=\Delta{\bm d}^{(E_u)}_1$ to the charge current. The $\zeta$ function is constrained by symmetries of the equilibrium order parameter (${\bm d}^{E_u}_1$) and bosonic field (${\bm d}^{\Gamma}_j$). In addition to the chirality mode ($\mathcal{D}^{-}_{E_u,2}$), long-lived massive bosons supported by sub-dominant pairing interactions ($\mathcal{D}^{-}_{A_{1u}}$ and $\mathcal{D}^{-}_{A_{2u}}$) are responsible for pronounced absorption peaks in the transverse EM response.
In Table~\ref{table1}, we summarize the coupling of $\mathcal{D}^{\rm C}_{\Gamma,j}$ to EM fields with the propagation vectors ${\bm Q}\parallel \hat{\bm z}$ and ${\bm Q}\perp \hat{\bm z}$ for the odd-parity ground-states ($A_{1u}$, $A_{2u}$, and $E_u$). 
For ${\bm Q}=\hat{\bm y}$ and ${\bm A}=\hat{\bm x}$, the tensor $v^{\mu}_{\rm F}v^{\nu}_{\rm F}$ in Eq.~\eqref{eq:zeta} reduces to $v^{x}_{\rm F}v^{y}_{\rm F}\sim p_xp_y$. As $\bar{\lambda}$ is an even function on ${\bm p}$, only the chiral Higgs mode with ${\bm d}^{(E_u)}_2$ couples to the transverse EM field. 
The selection rules for the ground-states, $A_{1u}$ and $A_{2u}$, are obtained by replacing ${\bm d}^{E_u}_1$ to ${\bm d}^{A_{1u}}$ and ${\bm d}^{A_{2u}}$ in Eq.~\eqref{eq:zeta}, respectively. The contributions from the $A_{1u}$ state are prohibited by the enlarged symmetry ${\tt D_{\infty}}$ around the small pocket of the Fermi surface. The breaking of ${\tt D_{\infty}}\rightarrow\d3d$ lifts this super-selection rule. In addition, the coupling of the nematicity mode to the charge current is prohibited by the particle-hole symmetry. 
%

\begin{table}[t!]
\caption{\label{table1}Selection rules for the coupling of transverse EM waves with ${\bm Q}$ to the bosonic modes, $\mathcal{D}^-_{\Gamma,j}$ (third-to-sixth columns). The second column denotes the irreducible representations of the ground-state (G.S.) order parameter. We take $\hat{\bm z}$ along the (111) axis of the ${\tt D_{3d}}$ crystal.} 

\begin{ruledtabular}
\begin{tabular}{cccccccc}
& ${\bm Q}$ & G.S. & $\mathcal{D}^-_{A_{1u}}$ & $\mathcal{D}^-_{A_{2u}}$ & $\mathcal{D}^-_{E_{u},1}$ & $\mathcal{D}^-_{E_{u},2}$ & \\
\hline
&${\bm Q}\parallel \hat{\bm z}$ &
\begin{tabular}{c}
$A_{1u}$ \\ $A_{2u}$ \\ $E_{u,1}$ \\ $E_{u,2}$ 
\end{tabular}
&
\begin{tabular}{c}
--- \\ --- \\ --- \\ --- 
\end{tabular}
&
\begin{tabular}{c}
--- \\ --- \\ ${\bm A}\perp \hat{\bm z}$ \\ ${\bm A}\perp \hat{\bm z}$
\end{tabular}
&
\begin{tabular}{c}
--- \\ ${\bm A}\perp \hat{\bm z}$ \\ --- \\ ---
\end{tabular}
&
\begin{tabular}{c}
--- \\ ${\bm A}\perp \hat{\bm z}$ \\ --- \\ --- 
\end{tabular}
& \\
\hline
&${\bm Q}\perp \hat{\bm z}$ &
\begin{tabular}{c}
$A_{1u}$ \\ $A_{2u}$ \\ $E_{u,1}$ \\ $E_{u,2}$ 
\end{tabular}
&
\begin{tabular}{c}
--- \\ --- \\ --- \\ --- 
\end{tabular}
&
\begin{tabular}{c}
--- \\ --- \\ ${\bm A}\parallel \hat{\bm z}$ \\ ${\bm A}\parallel \hat{\bm z}$
\end{tabular}
&
\begin{tabular}{c}
--- \\ ${\bm A}\parallel \hat{\bm z}$ \\ --- \\ ${\bm A}\perp \hat{\bm z}$
\end{tabular}
&
\begin{tabular}{c}
--- \\ ${\bm A}\parallel \hat{\bm z}$ \\ ${\bm A}\perp \hat{\bm z}$ \\ --- 
\end{tabular}
& \\
\end{tabular}
\end{ruledtabular}
\end{table}

\begin{figure}[t!]
\includegraphics[width=80mm]{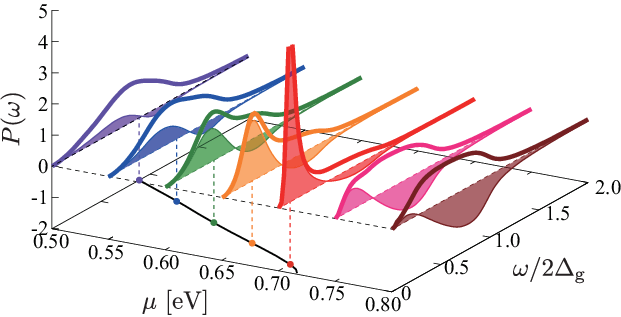}
\caption{Power absorption, $P(\omega)$, in the nematic state with the nematic angle $\vartheta=0$ for $T=0.05T_{\rm c}$, where we set ${\bm A}\parallel {\bm x}$ and ${\bm Q}\parallel{\bm y}$ and all parameters are same as those in Fig.~\ref{fig:mass}. The shaded area {shows} the contribution of {the} bosonic excitations, $P^{\rm CM}(\omega)$. The mass gaps of the chirality mode, $M^-_{E_u,2}$, are also shown.}
\label{fig:power}
\end{figure}

Figure~\ref{fig:power} shows the power absorption, $P(\omega)$, and the bosonic excitation contribution, $P^{\rm CM}$, for the $E_{u,1}$ nematic ground-state at $T=0.05T_{\rm c}$ {for} ${\bm Q}\parallel \hat{\bm y}$ and ${\bm A}\parallel\hat{\bm x}$. According to the selection rules, the EM field couples to only the chiral Higgs mode, $\mathcal{D}^-_{E_u,2}$. For $\mu = 0.5$eV, a broad peak in the spectrum appears around $\omega = 2\Delta _{\rm g}$. This broad peak arises primarily from the continuum of Bogoliubov QPs, i.e., $P^{\rm QP}$, and to a lesser extent the chiral Higgs mode, consistent with the large damping rate of the chiral Higgs mode shown in Fig.~S3 in Ref.~\cite{SM}.
As $\mu$ further increases the broad peak sharpens and shifts to lower frequencies. The pronounced peak originates from resonant absorption of the EM field by the chiral Higgs mode. The shift to lower frequency reflects the softening of the mass gap of these modes. Hence, the precursor to the dynamical instability of the nematic state to the chiral state is captured as a pronounced low-frequency peak in the EM power absorption.

Transverse EM fields with different configurations of ${\bm A}({\bm Q})$ couple to different bosonic modes. For instance, the EM field with ${\bm A}\parallel \hat{\bm z}$ and ${\bm Q}\parallel {\bm x}$ couples to the chiral $A_{2u}$ mode, ${\bm d}(t)=\Delta {\bm d}^{\rm E_u}_1+i\epsilon(t){\bm d}^{A_{2u}}$. Similarly to Fig.~\ref{fig:power}, a pronounced low-frequency peak appears in $P(\omega)$ as a consequence of the resonant contribution of the chiral $A_{2u}$ mode (see Fig.~S4 in Ref.~\cite{SM}).

{\it Signature of nematicity mode.} Finally, we note that the nematicity mode makes a significant contribution to dynamical spin susceptibility, $\chi _{zz}({\bm Q},\omega)=\chi^{\rm QP}_{zz}({\bm Q},\omega)+\chi^{\rm CM}_{zz}({\bm Q},\omega)$, where $\chi^{\rm QP}_{zz}({\bm 0},0)$ is the spin susceptibility of the equilibrium nematic state and $\chi^{\rm CM}_{zz}$ corresponds to the response of the nematicity mode (see Sec.~S4 in Ref.~\cite{SM}). As shown in Fig.~\ref{fig:mass}(b), the mass of the nematicity mode is sensitive to $T^{(Eu,2)}_{\rm c}/T^{(E_u,1)}_{\rm c}$, i.e., weak symmetry-breaking perturbations to the ${\tt D_{\infty}}$. The resulting small mass gap is detected as a pronounced peak in $\chi^{\rm CM}_{zz}$ at rf-frequencies resonant with the mass gap. Therefore, dynamical susceptibility measurements may provide a probe for the intrinsic mechanism of pinning the nematic order. 

{\it Summary.} We have discovered theoretically two characteristic bosonic excitations in nematic SCs: nematicity and chirality modes. The Fermi surface evolution softens the mass gap of the chiral Higgs mode, and the mass shift reflects a distance from the nematic-to-chiral transition in low temperatures. We have also demonstrated that owing to the selection rule, only EM waves with ${\bm Q}\perp{\bm A}\perp\hat{\bm z}$ can directly couple to the chiral Higgs mode in nematic SCs. These results show that a pronounced peak observed in absorption measurements can be a direct probe for the chirality excitation energy from the nematic ground state.

Low-lying bosons ubiquitously exist in multi-component SCs with unconventional symmetry breaking, and the selection rule for their EM/magnetic responses is based on the generic argument with the particle-hole symmetry and gap/crystalline symmetries. Hence, EM response provides a spectroscopy of spontaneously broken symmetries and sub-dominant pairing interactions in the broad family of nematic SCs~\cite{machidaJPSJ18,andersenPRB18,casn3} and unconventional SCs.

\begin{acknowledgments} 
This work was supported by a Grant-in-Aid for Scientific Research on Innovative Areas ``Topological Materials Science'' (Grants No.~JP15H05852, No.~JP15H05855, and No.~JP15K21717) and ``J-Physics'' (JP18H04318) and JSPS KAKENHI (Grant No.~JP16K05448 and No.~JP17K05517). The research of J.A.S. was supported by the National Science Foundation Grant DMR-1508730, and in part by the Aspen Center for Physics, which is supported by National Science Foundation grant PHY-1607611. 
\end{acknowledgments} 

\newpage

\renewcommand{\thesection}{S\arabic{section}}
\renewcommand{\theequation}{S\arabic{equation}}
\setcounter{equation}{0}
\renewcommand{\thefigure}{S\arabic{figure}}
\setcounter{figure}{0}

\centerline{{\bfseries Appendix -- Supplementary Material}}

\section{S1. Quasiclassical transport theory}

Here we derive the solutions of the transport equations for the quasiclassical propagator, $\check{\mathfrak{g}}$. The propagator is obtained from the Green's function in Keldysh space, 
\beq
\check{G} = \begin{pmatrix}
\hat{G}^{\rm R} & \hat{G}^{\rm K} \\ 0 & \hat{G}^{\rm A}
\end{pmatrix},
\eeq
where $\hat{G}^{\rm R}$, $\hat{G}^{\rm A}$, and $\hat{G}^{\rm K}$ are the retarded, advanced, and Keldysh Green's functions in Nambu space, respectively.
A key feature of the quasiclassical approximation is that $\check{G}$ is sharply peaked at the Fermi surface embedded in the conduction band, and depends weakly on energies far away from it. We use this assumption to split the propagator into low and high energy parts, 
$\check{G} = \check{G}^{\rm low} + \check{G}^{\rm high}$, where
$\check{G}^{\rm low} (p,Q) = \check{G}(p,Q)$ 
for $|\varepsilon| < \varepsilon_{\rm c}$ and otherwise $\check{G}^{\rm low} (p,Q) =0$ ($p$ and $Q$ are the relative and center-of-mass momentum, respectively)~\cite{rai95}.
The cutoff energy, $\varepsilon _{\rm c}$, is taken to be $T_{\rm c} \ll \varepsilon_{\rm c}\ll E_{\rm F}$ ($T_{\rm c}$ is the transition temperature). The high-energy part of this propagator renormalizes bare interactions to effective interactions parametrized by phenomenological parameters, such as effective coupling constant and Landau Fermi liquid parameters. The low-energy propagator defines the quasiclassical Keldysh propagators, $\hat{\mathfrak{g}}^{\rm x}$ (${\rm x}={\rm R}, {\rm A}, {\rm K}$) as an integral over a low-energy, long-wavelength shell, $|{\bm v}_{\rm F}\cdot({\bm p}-{\bm p}_{\rm F})|$, in momentum space near the Fermi surface,
\begin{align}
\hat{\mathfrak{g}}^{\rm x}({\bm p}_{\rm F},\varepsilon,Q) 
=& \frac{1}{a}\int^{\varepsilon_{\rm c}}_{-\varepsilon_{\rm c}} d\xi_p
\hat{\tau} _3 \hat{G}^{\rm x}({\bm p},\varepsilon , Q) \nn \\
=& \begin{pmatrix}
g^{\rm x}_0 + {\bm g}^{\rm x}\cdot {\bm \sigma} & i\sigma _2 f^{\rm x}_0 + i({\bm \sigma}\cdot{\bm f}^{\rm x})\sigma _2 \\
i\sigma _2 \bar{f}^{\rm x}_0 + i\sigma _2 ({\bm \sigma}\cdot \bar{\bm f}^{\rm x})
& \bar{g}^{\rm x}_0+\bar{\bm g}^{\rm x}\cdot {\bm \sigma}^{\rm T}
\end{pmatrix},
\end{align}
where $\xi _p = {\bm v}_{\rm F}\cdot({\bm p}-{\bm p}_{\rm F})$ is the quasiparticle excitation energy, $a$ is the spectral weight of the low-energy quasiparticle resonance and ${\bm \sigma}^{\rm T}$ is the transpose of the spin Pauli matrix ${\bm \sigma}$. We also introduce the abbreviation $Q\equiv ({\bm Q},\omega)$. The quasiclassical propagatoris governed by Eilenberger's transport  equation~\cite{eil68}, 
\begin{align}
\left[
\varepsilon\hat{\tau} _3 - \check{\mathfrak{h}}({\bm p}_{\rm F},Q),  \check{\mathfrak{g}}({\bm p}_{\rm F},\varepsilon,Q)
\right]_{\circ} =\eta\check{\mathfrak{g}}({\bm p}_{\rm F},\varepsilon,Q),
\label{eq:trans2}
\end{align}
where $\eta = {\bm v}_{\bm F}\cdot{\bm Q}$.
$\check{\mathfrak{h}}$, is the quasiclassical self-energy matrix in Keldysh space, where $\hat{\mathfrak{h}}^{\rm K}=0$ and $\hat{\mathfrak{h}}^{\rm R}=\hat{\mathfrak{h}}^{\rm A}\equiv\hat{\mathfrak{h}}$ is given as
\beq
\hat{\mathfrak{h}}
= \hat{\Sigma}+\hat{\Delta}
= 
\begin{pmatrix}
\Sigma _0 + {\bm \Sigma}\cdot {\bm \sigma} & i\sigma _{\mu}\sigma _2 d_{\mu} \\
i\sigma _2\sigma _{\mu}\bar{d}_{\mu}
& \bar{\Sigma}_0+\bar{\bm \Sigma}\cdot {\bm \sigma}^{\rm T}
\end{pmatrix}.
\eeq
The term, $\hat{\Sigma}$, contains the coupling to an electromagnetic field, and $\hat{\Delta}$ represents the off-diagonal pairing self energy, or order parameter. The transport equation and self energies are supplemented by Eilenberger's normalization condition $\check{\mathfrak{g}}\circ \check{\mathfrak{g}} = -\pi^2$ (for the notation, see the main text).

Instead of directly solving the Keldysh transport equation~\eqref{eq:trans2}, we derive the Keldysh propagator from the Matsubara propagator by analytic continuation to the real energy axes , e.g. $i\varepsilon _n\rightarrow\varepsilon + i0^{+}$ followed by $i\omega_m\rightarrow\omega + i0^{+}$~\cite{saulsPRB17}. Thus,
\beq
T\sum_{\varepsilon_n}\,\mathfrak{g}^{\rm M}(\varepsilon_n;\omega_m)
\xrightarrow[i\omega_m\rightarrow\omega+i0^{+}]{}
\int_{-\varepsilon _{\rm c}}^{+{\varepsilon _{\rm c}}}\,\frac{d\varepsilon}{4\pi i}
\mathfrak{g}^{\rm K}(\varepsilon;\omega)
\,.
\label{eq:analytic}
\eeq
To calculate the Keldysh propagator, $\hat{\mathfrak{g}}^{\rm K}$, we generalize the Matsubara transport equation for the two-time/frequency non-equilibrium Matsubara propagator~\cite{saulsPRB17},
\beq
\left[i\varepsilon \hat{\tau} _3 - \hat{\mathfrak{h}} \right]\circ\hat{\mathfrak{g}}^{\rm M}
- \hat{\mathfrak{g}}^{\rm M}\circ\left[i\varepsilon\hat{\tau}_3 - \hat{\mathfrak{h}}\right] 
-\eta\hat{\mathfrak{g}}^{\rm M}= 0
\eeq
where the 
$A\circ B(\varepsilon _{n_1},\varepsilon _{n_2})\equiv T\sum _{n_3}\,
 A(\varepsilon _{n_1},\varepsilon _{n_3})\,B(\varepsilon _{n_3},\varepsilon _{n_2})$ 
is a convolution in Matsubara energies.
For the two-frequency propagator,the normalization condition is also a convolution
product in Matsubara frequencies,
\beq
\hat{\mathfrak{g}}^{\rm M}\circ\hat{\mathfrak{g}}^{\rm M}
\equiv
T\sum_{\varepsilon_{n_3}}
\hat{\mathfrak{g}}^{\rm M}(\varepsilon_{n_1},\varepsilon_{n_3})
\hat{\mathfrak{g}}^{\rm M}(\varepsilon_{n_3},\varepsilon_{n_2})
=
-\frac{\pi^2}{T}\delta _{\varepsilon_{n_1},\varepsilon_{n_2}}.
\eeq
We now express the full propagator as the sum of the equilibrium propagator and a non-equilibrium correction,
\begin{align}
\hat{\mathfrak{g}}^{\rm M}({\bm p}_{\rm F},{\bm Q};\varepsilon_{n_1},\varepsilon_{n_2})
=
\hat{\mathfrak{g}}^{\rm M}_{0}({\bm p}_{\rm F},\varepsilon_{n_1}) \,\frac{1}{T}\delta _{\varepsilon_{n_1},\varepsilon_{n_2}}
+
\delta\hat{\mathfrak{g}}^{\rm M}({\bm p}_{\rm F},{\bm Q};\varepsilon_{n_1},\varepsilon_{n_2}).
\label{eq:linearg}
\end{align}

\subsection{Model Hamiltonian and basis functions}

The parent materials of carrier-doped topological insulators, $M_x$Bi$_2$Se$_3$ ($M={\rm Cu}, {\rm Sr}, {\rm Nb}$), are composed of spin-$1/2$ fermions with orbital degrees of freedom. The low-energy structure is governed by two $p_z$ orbitals localized on the lower and upper sides of the quintuple layer. The $4\times 4$ effective Hamiltonian, $\xi ({\bm p})$, relevant to the parent material is given by 
\begin{align}
\xi ({\bm p}) = c({\bm p}) + m({\bm p})\sigma _x + v_z f(p_z)\sigma _y 
+ v\left( {\bm p}\times {\bm s} \right)_z\sigma _z ,
\label{eq:htik2}
\end{align}
where $c ({\bm p}) = c_0 + c_1 f_{\perp}({\bm p})+c_2p^2_{\parallel}$, $m ({\bm p}) = m_0 + m_1 f_{\perp}({\bm p}) + m_2p^2_{\parallel}$, and $\mu$ are the diagonal self-energy correction, band gap, and the chemical potential, respectively ($p^2_{\parallel}\equiv p^2_x+p^2_y$). Nearest-neighbor hopping along the $z$ direction gives
$f_z(p_z)=\frac{1}{c} \sin(p_zc)$ and $f_{\perp} =\frac{2}{c^2}[1-\cos(p_zc)]$.
We have also introduced the spin and orbital Pauli matrices, ${ s}_{\mu}$ and ${ \sigma}_{\mu}$ ($\mu=x,y,z$). We take $z$-axis along the (111) direction of the crystal along which the quintuple layers are stacked by van der Waals gap. The effective Hamiltonian approximately holds the ${\rm D}_{\infty}$ including the ${\rm SO}(2)_{J_z}$ symmetry about the $z$-axis, while the higher order correction on $p$ introduces the three mirror planes and threefold rotational symmetry in the $xy$ plane.

In this work, we consider the linear response of nematic superconductors to electromagnetic fields. The vector potential ${\bm A}$ is introduced in Eq.~\eqref{eq:htik2} by Peierls substituion, ${\bm p}\rightarrow {\bm p}-e{\bm A}$. In addition, the Zeeman term in the parent topological insulator is given by adding the following term in Eq.~\eqref{eq:htik2}
\beq
\mathcal{H}_{\rm z} = \sum _{\mu,i} \frac{1}{2}g_{i\mu} \mu _{\rm B} s _i H_i\sigma _{\mu},
\label{eq:hz}
\eeq
where $\mu _{\rm B}$ is the Bohr magneton and $H_i$ is the $i$th component of the Zeeman field ($i=x,y,z$). The $g$-factor of the parent topological insulator is given by $g_{i\mu}$ ($\mu = 0, x, y,z$). For Bi$_2$Se$_3$, $g_{x0}=g_{y0}=-8.92$, $g_{z0}=-21.3$, $g_{xx}=g_{yx}=0.68$, and $g_{zx}=-29.5$ and $g_{i\mu}=0$ otherwise~\cite{LiuPRB10,hashimotoJPSJ13}.

Let us now introduce the gap functions belonging to the irreducible representations of the ${\rm D}_{\rm 3d}$ crystalline symmetry of the compounds $M_x$Bi$_2$Se$_3$. The $4\times 4$ matrix form of the superconducting gap function is given by 
\beq
\Delta({\bm R})  = \sum^{n_{\Gamma}}_{j=1}\eta^{(\Gamma)}_j({\bm R})W^{(\Gamma)}_{j}is_y,
\label{eq:eu}
\eeq
where $\Gamma = A_{1g}, A_{1u}, A_{2u},E_u$ are the even-parity and odd-parity irreducible representations of ${\rm D}_{\rm 3d}$ with the dimension $n_{\Gamma}$ and the basis functions $\{ W^{(\Gamma)}_{1},\cdots, W^{(\Gamma)}_{n_{\Gamma}}\}$. The $4\times 4$ matrix $W^{(\Gamma)}_{j}$ is given as 
$W^{(A_{1g})} = \{ 1, \sigma_x\}$
for the even parity state, and 
$W^{(A_{1u})} = \sigma _ys_z$, $W^{(A_{2u})} = \sigma_z$, 
and $(W^{(E_u)}_1, W^{(E_u)}_2) = (\sigma _ys_x, \sigma_ys_y)$ for the odd-parity states.

The Hamiltonian in Eq.~\eqref{eq:htik2} is diagonalized as $U^{\dag}({\bm p})\xi ({\bm p})U({\bm p})={\rm diag}[E_{\rm CB}({\bm p}), E_{\rm VB}({\bm p})]$. The conduction band energy, $E _{\rm CB}({\bm p})=c-\mu+\sqrt{m^2+v^zf^2+v^2p^2_{\parallel}}$, is separated from the valence band, $E_{\rm VB}({\bm p})=c-\mu-\sqrt{m^2+v^zf^2+v^2p^2_{\parallel}}$, by the band gap $2|m_0|$ at the $\Gamma$ point. The intercalation of Cu, Sr, and Nb atoms into the van der Waals gap increases the carrier density of the conduction band and generates a small electron Fermi pocket around the $\Gamma$ point. Since the band gap is $m_0 \!=\! -0.28~{\rm eV}$ and $\mu$ is the same order as $|m_0|$, both energy scales are much larger than the superconducting gap. Hence, it is natural to employ the quasiclassical approximation which takes account of only the electron states in the conduction band.

Let $\mathcal{P}_{\rm CB}$ be a projection operator onto the conduction band. The pair potential projected onto the conduction band is parameterized with  the even-parity scalar field $\psi({\bm p})=\psi(-{\bm p})$ and odd-parity ${\bm d}$-vector field ${\bm d}({\bm p})=-{\bm d}(-{\bm p})$ as 
\beq
\mathcal{P}_{\rm CB}[U^{\dag}({\bm p})\Delta U^{\ast}({\bm p})] = {\psi}({\bm p})is_y + is_{\mu}s_yd_{\mu}({\bm p}).
\eeq
The repeated Greek indices imply the sum over the vector components of the spin $S=1$ basis, $x,y,z$, constructed to provide bases of the irreducible representation $\Gamma$. $V^{(\Gamma)}_i>0$ is the coupling constant for the representation $\Gamma$.
In the band representation~\cite{hashimotoJPSJ13}, the $E_u$ state has $\psi^{(E_u)}_{1,2}=0$ and
\begin{gather}
{\bm d}^{(E_u)}_{1}
= \left[\left(
1-\mathcal{E}\frac{p^2_y}{p^2_{\parallel}}
\right)\frac{v_zp_z}{\tilde{\varepsilon}(p_z)}, 
\mathcal{E}
\frac{p_xp_y}{p^2_{\parallel}}\frac{v_zp_z}{\tilde{\varepsilon}(p_z)}, 
 -\frac{m}{\tilde{\varepsilon}}\frac{vk_x}{\tilde{\varepsilon}(p_z)} 
\right], \\
{\bm d}^{(E_u)}_{2}
= \left[ \mathcal{E}
\frac{p_xp_y}{p^2_{\parallel}}\frac{v_zp_z}{\tilde{\varepsilon}(p_z)}, 
\left(
1-\mathcal{E}\frac{p^2_x}{p^2_{\parallel}}
\right)\frac{v_zp_z}{\tilde{\varepsilon}(p_z)}, 
-\frac{m}{\tilde{\varepsilon}}\frac{vk_y}{\tilde{\varepsilon}(p_z)}
\right],
\end{gather}
where we set $\tilde{\varepsilon}\equiv\sqrt{m^2+v^zf^2+v^2p^2_{\parallel}}$, $\tilde{\varepsilon}(p_z)\equiv\tilde{\varepsilon}(0,0,p_z)$, and $\mathcal{E}\equiv
1-\tilde{\varepsilon}(p_z)/\tilde{\varepsilon}({\bm p})$. In the same manner, the basis functions in the band representation are given by 
\beq
\psi^{(A_{1g})}_1 ({\bm p})= \left\{ 1,~ m({\bm p})/\tilde{\varepsilon}({\bm p}) \right\}, 
\eeq
and ${\bm d}^{(A_{1g})}_1 ({\bm p})={\bm 0}$ for $A_{1g}$, 
\beq
{\bm d}^{(A_{1u})}_1 ({\bm p})=
\frac{1}{\tilde{\varepsilon}(p_z)}\left(
\frac{m({\bm p})}{\tilde{\varepsilon}({\bm p})}vp_x,
\frac{m({\bm p})}{\tilde{\varepsilon}({\bm p})}vp_y, v_zp_z
\right), 
\eeq
and $\psi^{(A_{1u})}_1 ({\bm p})= 0$ for $A_{1u}$, and 
\beq
{\bm d}^{(A_{2u})}_1 ({\bm p})=
\left(
- vp_y,vp_x, 0
\right)/\tilde{\varepsilon}({\bm p}),
\eeq
and $\psi^{(A_{2u})}_1 ({\bm p})= 0$ for $A_{2u}$.

\subsection{Nematic-to-chiral phase transition: Ginzburg-Landau theory}

Using the Ginzburg-Landau (GL) theory, we first show that for pairing governed by the $E_u$ representation, a nematic-to-chiral phase transition occurs at a critical chemical potential. Consider the ground state of the $E_u$ representation with the order parameter
\beq
{\bm d}({\bm p}_{\rm F}) = \eta _1 {\bm d}^{E_u}_1 ({\bm p}_{\rm F})
+ \eta _2 {\bm d}^{E_u}_2 ({\bm p}_{\rm F}).
\label{eq:dEu}
\eeq
The GL free energy up to fourth order is then given by 
\beq
\mathcal{F} = \alpha |{\bm \eta}|^2  + \beta _1 |{\bm \eta}|^4
+ \beta _2 |\eta _1 \eta^{\ast}_2 - \eta^{\ast}_1\eta_2|^2.
\label{eq:f}
\eeq
The thermodynamic stability of superconducting states below $T_{\rm c}$ requires $\alpha \propto T-T_{\rm c}$ and $\beta _1 > 0$. The fourth-order coefficient defined as 
\beq
\beta _2 = -\langle ({\bm d}_1\cdot{\bm d}_2)^2\rangle _{\rm FS}
+ \langle ({\bm d}_1\times {\bm d}_2)^2 \rangle _{\rm FS},
\eeq
determines the order parameter configuration ${\bm \eta}=(\eta _1,\eta_2)$, where $\langle ... \rangle _{\rm FS} \equiv\int dS_{\bm p} ... $ is an average over the Fermi surface that satisfies $\int dS_{\bm p} = 1$.

For $\beta _2 > 0$, the nematic state with $(\eta _1,\eta_2) = \Delta (\cos\vartheta,\sin\vartheta)$ is stable as the highly degenerate minima of $\mathcal{F}$ with respect to $\vartheta \in [0,\pi/2]$.
The gap structure has two point nodes in the $xy$ plane [Fig.~1(a) in the main text]. The continuous degeneracy with respect to $\vartheta$ is accidental, and is lifted by the sixth-order term representing the hexagonal warping of the Fermi surface, 
\beq
\mathcal{F}_6 = \kappa[(\eta^{\ast}_+ \eta _-)^3 + (\eta _+ \eta^{\ast}_-)^3],
\eeq 
with $\eta _{\pm} \equiv \eta _1\pm i\eta _2$, which pins the nematic angle $\vartheta$ to one of three equivalent crystal axes.

The $\beta _2 < 0$ region is covered by the chiral state, $(\eta _1, \eta _2) =\Delta(1,\pm i)$, which breaks time reversal symmetry. As a result of spin-orbit coupling, the chiral state, ${\bm d}_1 \pm i{\bm d}_2$, is also a nonunitary state, with two distinct gaps: fully gapped and a gap with point nodes at ${\bm p}=\pm{p}_{{\rm F},z}\hat{\bm z}$.

In $M_x$Bi$_2$Se$_3$, the intercalation of $M$ atoms between the quintuple layers modifies the $c$-axis length of the crystal, namely, the hopping parameters along the $z$-axis $(c_1, m_1, v_z)$. To incorporate the Fermi surface evolution, we follow Ref.~\cite{hashimoto14}: the set of parameters in Ref.~\cite{hashimotoJPSJ13} for $\mu = 0.4$eV and the half-value of $(c_1, m_1, v_z)$ for $\mu = 0.65$eV. The parameters for arbitrary $\mu$ are given by interpolating $(c_1, m_1, v_z)$ linearly with respect to $\mu$ (for the further information, see the next subsection). With this parametrization, the Fermi surface is opened along the $z$-axis for $\mu \gtrsim 0.5$eV. Using this set of parameters, we calculate the $\beta _2$ as a function of $\mu$. We find that there exists the critical value $\mu _{\rm c}=0.7$eV at which $\beta _2 = 0$ corresponding to a nematic-to-chiral phase transition.

\subsection{Nematicity and chirality modes: Time-dependent Ginzburg-Landau theory}

To understand characteristic bosonic excitations in the nematic ground state, we first solve the time-dependent Ginzburg-Landau (TDGL) theory. Let ${\bm \eta}(t)$ be a dynamical bosonic field representing the $E_u$ order parameter. The equation of motion for ${\bm \eta}(t)$ is obtained from the effective Lagrangian, 
\beq
\mathcal{L} \equiv \tau |\partial _t {\bm \eta}|^2 - \mathcal{F}[\eta _j,\eta^{\ast}_j ]
- \mathcal{F}_6[\eta _j,\eta^{\ast}_j ],
\label{eq:lagrange}
\eeq
with an effective inertia of Cooper pair fluctuations, $\tau > 0$. Although the effective Lagrangian formalism does not incorporate the contribution of Bogoliubov quasiparticles, it can quantitatively describe all the collective modes in the bulk superfluid $^3$He-B~\cite{saulsPRB17,mizushima18}. This also gives a tractable way to capture the bosonic excitations in unconventional SCs.
We then introduce the linear fluctuation of the order parameter, $\delta{\bm \eta} (t)\equiv {\bm \eta} (t) - {\bm \eta}$, in terms of two orthogonal nematic vectors, ${\bm b}_1=(\cos\vartheta,\sin\vartheta)$ and ${\bm b}_2=(-\sin\vartheta,\cos\vartheta)$, as
\beq
\delta {\bm \eta}^{\rm C}(t) 
=  \mathcal{D}^{\rm C}_1(t) {\bm b}_1 + \mathcal{D}^{\rm C}_2(t) {\bm b}_2.
\label{eq:deltad2}
\eeq
All the collective modes are separated to the four sectors. Two of them are in the ground state sector $\mathcal{D}^{\rm C}_1$ and the others are in the orthogonal sector $\mathcal{D}^{\rm C}_2$, where 
\beq
\mathcal{D}^{\rm C}_j=\mathcal{D}_j+{\rm C}\mathcal{D}^{\ast}_j
\eeq
is the parity eigenstate under particle-hole conversion.

Consider the phase fluctuation of the equilibrium sector, ${\bm d}(t) = \Delta e^{i\delta\theta}{\bm d}^{E_u}_1\approx \Delta (1+i\delta\theta){\bm d}^{E_u}_1$. Then, the mode with $\mathcal{D}^-_{1}=i\Delta \delta \theta$ corresponds to the NG mode associated with the broken $\gauge$ symmetry, which is gapped out by the Anderson-Higgs mechanism. The amplitude fluctuation is represented by $\mathcal{D}^+_{1} = \delta\epsilon(t)$, where ${\bm d}(t)=[\Delta+\delta \epsilon(t)]{\bm d}^{E_u}_1$. The fluctuation modes of the orthogonal basis to the equilibrium basis are represented by $\mathcal{D}^{\rm C}_{2}$. Let us suppose $\mathcal{D}^+_{2} = \mathcal{D}_{2} + \mathcal{D}^{\ast}_{2} = \Delta \delta \vartheta(t)$ and $\mathcal{D}^-_{2}=\mathcal{D}_{2}-\mathcal{D}^{\ast}_{2}=i\epsilon(t) \Delta$. Using these parameterizations, the $\mathcal{D}^+_{2}$ mode leads to ${\bm d}(t)=\Delta[{\bm d}^{E_u}_1+\delta\vartheta (t){\bm d}^{E_u}_2]$, then $[\eta_1(t),\eta_2(t)] = [1, \delta \vartheta(t)] \approx [\cos(\delta\vartheta(t)), \sin(\delta \vartheta(t))]$ for $\delta \theta\ll 1$. This is the pseudo-NG mode associated with the broken rotation symmetry or the fluctuation of the nematic order. The $\mathcal{D}^-_{2}$ mode gives rise to the fluctuation of the chirality or orbital angular momentum of Cooper pairs, ${\bm d}(t)= \Delta[{\bm d}^{E_u}_1+i\epsilon(t) {\bm d}^{E_u}_2]$. We term $\mathcal{D}^+_{2}$ and $\mathcal{D}^-_{2}$ the nematicity mode and chiral Higgs mode, respectively.

The Euler-Lagrange equation is obtained from $\mathcal{L}$ in Eq.~\eqref{eq:lagrange} as
\beq
-\partial^2_t \mathcal{D}^{\rm C}_j = \frac{1}{\tau} \frac{\delta^2\mathcal{F}_{\rm GL}}{\delta \mathcal{D}^{\rm C}_j \delta \mathcal{D}^{\rm C}_j} \equiv (M^{\rm C}_{j})^2\mathcal{D}^{\rm C}_j. 
\eeq
The mass gap of each bosonic mode, $M^{\rm C}_{j}$, represents the local curvature of $\mathcal{F}_{\rm GL}\equiv\mathcal{F}+\mathcal{F}_6$ around the GL equilibrium solution. 
The mass gap of the chiral Higgs mode is given by
\beq
M^-_{E_u,2} = \Delta \sqrt{2(\beta _2+3\kappa \Delta^2)/\tau}.
\label{eq:mass-}
\eeq
The nematicity mode, which is the NG mode associated with the nematic order, is gapped out by explicit symmetry breaking due to the hexagonal warping effect as
\beq
M^+_{E_u,2} = 6\Delta^2\sqrt{2\kappa/\tau}.
\label{eq:mass+}
\eeq
For $\kappa =0$, the mass gap of the chiral Higgs mode vanishes at $\beta _2 = 0$, corresponding to $\mu_{\rm c}= 0.7$ eV. This softening of the chiral Higgs mode reflects that for $\beta_2 < 0$, i.e., $\mu > \mu_{\rm c}$, the GL functional exhibits negative curvature around nematic ground state, i.e., the dynamical instability of the nematic state towards the chiral state.

We note that the TDGL theory does not incorporate the contribution of Bogoliubov quasiparticles. As the nodal structure of the nematic gap function changes from the point node to line node as $\mu$ increases, the quasiparticle contributions to the mass shift and damping of bosons become significant in the vicinity of the nematic-to-chiral phase transition. Below, using the quasiclassical Keldysh theory, we examine the mass shift and damping rate of the low-lying bosonic excitations in the nematic ground state. We demonstrate that thermally excited Bogoliubov quasiparticles lead to the significant mass shift in high temperatures, while the TDGL theory can qualitatively capture the characteristic bosonic modes in nematic SCs.

\subsection{Nematic-to-chiral phase transition: Quasiclassical theory}


Here we describe the self-consistent equations and thermodynamic potential in terms of the equilibrium propagator, $\hat{\mathfrak{g}}^{\rm M}_{0}({\bm p}_{\rm F},\varepsilon_{n_1})$. 
%
The equilibrium propagator for unitary states, ${\bm d}\times {\bm d}^{\ast}={\bm 0}$, is
\beq
\hat{\mathfrak{g}}^{\rm M}_0({\bm p}_{\rm F},\varepsilon _n) = -\pi\frac{i\varepsilon _n\hat{\tau} _3 - \hat{\Delta}({\bm p}_{\rm F})}
{\sqrt{\varepsilon _n^2 + |{\bm d}({\bm p}_{\rm F})|^2}},
\label{eq:g0unitary}
\eeq
where 
\beq
\hat{\Delta}({\bm p}_{\rm F}) \equiv 
\begin{pmatrix}
0 & i\sigma _{\mu}\sigma _2 d_{\mu}({\bm p}_{\rm F}) \\
i\sigma _2\sigma _{\mu}{d}^{\ast}_{\mu}({\bm p}_{\rm F})
& 0
\end{pmatrix}.
\eeq
is the equilibrium superconducting order parameter matrix in the Nambu space. For non-unitary states, ${\bm q}\equiv i{\bm d}\times {\bm d}^{\ast}\neq{\bm 0}$, the spin-triplet components of the anomalous propagator are given by 
\beq
{\bm f}^{\rm M}({\bm p}_{\rm F},\varepsilon _n) = \frac{\sqrt{2\alpha}}{1-\alpha^2q^2}\left[
{\bm d}+i\alpha{\bm d}\times {\bm q} \right],
\label{eq:fM}
\eeq
where 
\begin{align}
\alpha ({\bm p}_{\rm F},\varepsilon_n) \equiv 
\frac{1}{\varepsilon^2_n+|{\bm d}({\bm p}_{\rm F})|^2+\sqrt{[\varepsilon^2_n+|{\bm d}({\bm p}_{\rm F})|^2]^2-[{\bm q}({\bm p}_{\rm F})]^2}}.
\end{align}
These propagators obey the normalization condition 
$
[\hat{\mathfrak{g}}^{\rm M}_{0}({\bm p}_{\rm F},\varepsilon_{n})]^2 = - \pi^2
$.

The pair potential, $(\eta_1,\eta_2)$, at the temperature $T$ is determined by solving the gap equation
\beq
d_{\mu}({\bm p}_{\rm F}) = T\sum _{|\varepsilon _n| < \varepsilon _{\rm n,c}} 
\left\langle
V_{\mu\nu}({\bm p}_{\rm F},{\bm p}^{\prime}_{\rm F}) f_{\nu}({\bm p}^{\prime}_{\rm F},\varepsilon_n)
\right\rangle^{\prime}_{\rm FS}
\eeq
where $\langle \cdots\rangle _{\rm FS}=\int dS_{\bm p}\cdots$ is an average over the Fermi surface that satisfies $\int dS_{\bm p}=1$.
In the nematic state, the gap equation is recast into
\beq
\frac{1}{V^{(E_u)}} = \pi T\sum _{|\varepsilon _n| < \varepsilon _{\rm n,c}}
\left\langle \frac{|{\bm d}^{(E_u)}_{1}({\bm p}_{\rm F})|^2}{\sqrt{\varepsilon _n^2 + |{\bm d}({\bm p}_{\rm F})|^2}}
\right\rangle _{\rm FS}.
\label{eq:gap0}
\eeq
We here assume the separable form of the pairing interaction 
\begin{align}
V_{\mu\nu}({\bm p}_{\rm F},{\bm p}^{\prime}_{\rm F})  = -\sum _{\Gamma}\sum^{n_{\Gamma}}_{j=1}
V^{(\Gamma)}_j {\bm d}^{(\Gamma)}_{\mu,j}({\bm p}_{\rm F}){\bm d}^{(\Gamma)\ast}_{\nu,j}({\bm p}_{\rm F}^{\prime}),
\end{align}
which comprises attractive interactions ($V^{(\Gamma)}_j>0$) in the irreducible representations of the symmetry group ${ D}_{\rm 3d}$, $\Gamma = \{ A_{1g}, A_{1u}, A_{2u}, E_{u}\}$. In calculating Eq.~\eqref{eq:gap0}, we utilize the fact that $\varepsilon _{\rm c}$ and $V^{(\Gamma)}$ are related to measurable quantity, the bulk transition temperature $T_{\rm c}$, by linearized gap equation
\begin{align}
\frac{1}{V^{(E_u)}_1} = \langle |{\bm d}^{(E_u)}_{1}({\bm p}_{\rm F})|^2\rangle_{\rm FS} K(T) ,
\end{align}
where $K(T)$ is the digamma function of argument $\varepsilon _{\rm c}/2\pi T \gg 1$,
\beq
K(T) = \pi T \sum_{|\varepsilon_n|<\varepsilon_{{\rm c}}}
\frac{1}{|\varepsilon_n|} \approx \ln \frac{1.13\varepsilon _{\rm c}}{T}.
\eeq
This relation can be utilized to eliminate $\varepsilon _{\rm c}$ and $V^{(E_u)}_1$ from the gap equation, and Eq.~\eqref{eq:gap0} reduces to 
\beq
\langle D \rangle_{\rm FS} \ln \frac{T}{T_{\rm c}}
= \pi T\sum _{n}
\left\langle \frac{D}{\sqrt{\varepsilon _n^2 + |{\bm d}|^2}}
- \frac{D}{|\varepsilon _n|}
\right\rangle _{\rm FS},
\label{eq:gapeq1}
\eeq
which is free from the ultraviolet divergence. In the same way, the gap equation for the non-unitary chiral state is given by
\begin{align}
\langle {D}\rangle_{\rm FS} \ln \frac{T}{T_{\rm c}}
&= \pi T\sum _{n}
\bigg\langle
\frac{\sqrt{2\alpha}}{1-\alpha^2{\bm q}^2} \nn \\
&\times \left\{
D + \alpha \Delta^2 \left( 
|{\bm d}^{E_u}_1\cdot{\bm d}^{E_u}_2|^2 - |{\bm d}^{E_u}_1|^2|{\bm d}^{E_u}_2|^2
\right)
\right\}- \frac{D}{|\varepsilon _n|}
\bigg\rangle _{\rm FS}.
\label{eq:gapeq2}
\end{align}
Here we set $D\equiv |{\bm d}^{E_u}_1|^2$ for the nematic state and $D\equiv (|{\bm d}^{E_u}_1|^2+|{\bm d}^{E_u}_2|^2)/2$ for the chiral state. We solve the gap equations \eqref{eq:gapeq1} and \eqref{eq:gapeq2} to obtain the self-consistent solution of $\Delta(T)$ for the nematic and chiral states at a given $\mu$ and $T$.

To compute the phase diagram of superconducting topological insulators, we need to introduce a free energy functional in terms of the equilibrium quasiclassical propagators and self-consistent pair potentials.  Following Ref.~\onlinecite{vorontsovPRB03} and using the Luttinger-Ward functional formalism, we obtain the free energy functional relative to the normal state as
\begin{align}
\Delta \Omega =& N_{\rm F} \langle |{\bm d}|^2\rangle\ln\frac{T}{T_{\rm c}} \nn \\
&
- N_{\rm F}\int^{1}_0 d\lambda T\sum _n \left[ 
\left\langle {\bm d}\cdot\bar{\bm f}^{\rm M}_{\lambda} + {\bm d}^{\ast}\cdot{\bm f}^{\rm M}_{\lambda} 
\right\rangle 
-\frac{\langle |{\bm d}|^2\rangle}{|\varepsilon _n|^2}\right]
\end{align}
where ${\bm f}_{\lambda}$ is obtained from Eqs.~\eqref{eq:g0unitary} and \eqref{eq:fM} with replacing ${\bm d}$ to $\lambda{\bm d}$. 

\begin{figure}[b!]
\includegraphics[width=85mm]{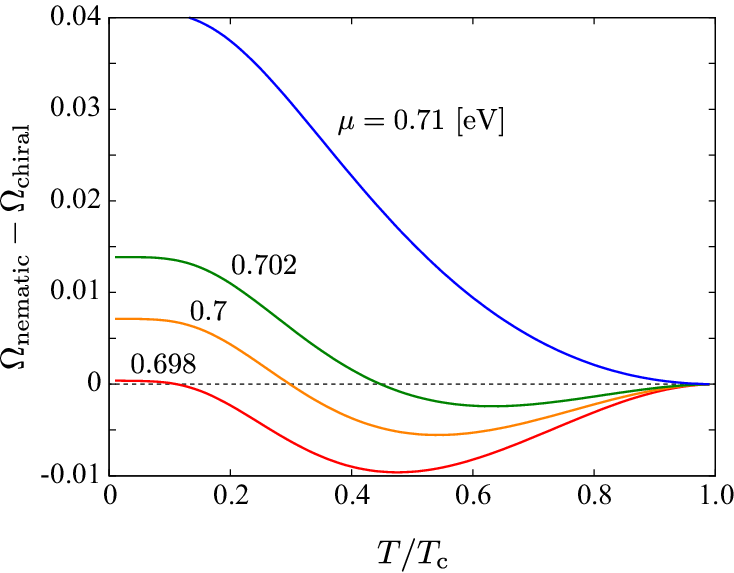}
\caption{Free energies of the nematic and chiral states for various $\mu$ in the vicinity of the nematic-chiral phase transition.}
\label{fig:s1}
\end{figure}

We first solve the gap equations \eqref{eq:gapeq1} and \eqref{eq:gapeq2} for the nematic and chiral state at $(T,\mu)$, respectively. Then we calculate the thermodynamic potential, $\Delta\Omega$, with the gap functions and the anomalous propagators and compute the phase diagram presented in the main text. Here we utilize the set of parameters for $\mu = 0.4$eV~\cite{hashimotoJPSJ13}: $m_0=-0.28~{\rm eV}$, $m_1/c^2=0.216~{\rm eV}$, $m_2/a^2=56.6~{\rm eV}$, $v_z/c=0.32~{\rm eV}$, and $v/a=0.56~{\rm eV}$, where $a=4.076\AA$ and $c=29.830\AA$ are the lattice constants of the parent material. For $\mu = 0.65$eV, we set the half-value of $(c_1, m_1, v_z)$, and the parameters for arbitrary $\mu$ are given by interpolating $(c_1, m_1, v_z)$ linearly with respect to $\mu$. The linear interpolation models the Fermi surface evolution of doped Bi$_2$Se$_3$, and the Fermi surface is opened along the $z$-axis for $\mu \gtrsim 0.5$eV. In Fig.~\ref{fig:s1}, we show the free energies of the nematic and chiral states in the superconducting topological insulators $M_x$Bi$_2$Se$_3$.

\subsection{Solutions of  linear response functions}

To linear order in $\delta\hat{\mathfrak{g}}^{\rm M}$ the normalization condition for the non-equilibrium correction to the propagator becomes
\beq
\hat{\mathfrak{g}}^{\rm M}_{0}(\varepsilon_n+\omega_m)\delta\hat{\mathfrak{g}}^{\rm M}(\varepsilon_n,\omega_m)
+
\delta\hat{\mathfrak{g}}^{\rm M}(\varepsilon_n,\omega_m)\hat{\mathfrak{g}}^{\rm M}_{0}(\varepsilon_n)
=0,
\label{eq:norm1}
\eeq
where we use $\varepsilon_n$ ($\omega_m$) for the Matsubara frequency for fermions (bosons). Thus, we set
$\varepsilon_{n_1} = \varepsilon_{n}+\omega_{m}$,
$\varepsilon_{n_2} = \varepsilon_{n}$, and
$\delta\hat{\mathfrak{g}}^{\rm M}({\bm p}_{\rm F},{\bm q};\varepsilon_{n_1},\varepsilon_{n_2})
\equiv
\delta\hat{\mathfrak{g}}^{\rm M}({\bm p}_{\rm F},{\bm q};\varepsilon_{n},\omega_{m})$.
Substituting Eq.~\eqref{eq:linearg} into Eq.~\eqref{eq:trans2}, the linearized transport equation for the non-equilibrium Matsubara propagator is given by
\begin{gather}
\left\{ i(\varepsilon _n + \omega _m )\hat{\tau}_3 - \hat{\Delta}({\bm p}_{\rm F})\right\}\delta \hat{\mathfrak{g}}^{\rm M}
-\delta \hat{\mathfrak{g}}^{\rm M}\left\{ i\varepsilon _n\hat{\tau}_3 
- \hat{\Delta}({\bm p}_{\rm F})\right\} -\eta \delta \hat{\mathfrak{g}}^{\rm M}  \nn \\
+ \hat{\mathfrak{g}}^{\rm M}_0(\varepsilon _n +\omega _m)\delta\hat{\mathfrak{h}}
-\delta \hat{\mathfrak{h}}\hat{\mathfrak{g}}^{\rm M}_0(\varepsilon _n) = 0,
\label{eq:eomM1}
\end{gather}
where $\delta\hat{\mathfrak{h}}\equiv \hat{\mathfrak{h}}-\hat{\Delta}$. For unitary states, we solve this equation using the normalization condition \eqref{eq:norm1} and the solutions for $\mathfrak{g}_0$ in Eq.~\eqref{eq:g0unitary}.
The linear response of the Matsubara propagator is then given by
\begin{align}
\delta \hat{\mathfrak{g}}^{\rm M} =& \frac{1}{D^2_++\eta^2}\bigg[
\frac{D_+}{\pi}\left\{
\hat{\mathfrak{g}}^{\rm M}_0(\varepsilon _n + \omega _m) \delta \hat{\mathfrak{h}} \hat{\mathfrak{g}}^{\rm M}_0(\varepsilon _n)
+ \pi^2\delta\hat{\mathfrak{h}}
\right\} \nn \\
&-\eta\left\{+
\delta \hat{\mathfrak{h}} \hat{\mathfrak{g}}^{\rm M}_0(\varepsilon _n) - \hat{\mathfrak{g}}^{\rm M}_0(\varepsilon _n + \omega_ m)\delta \hat{\mathfrak{h}}
\right\}
\bigg],
\end{align}
where 
$D({\bm p}_{\rm F},\varepsilon_n) = \sqrt{\varepsilon^2_n + |{\bm d}({\bm p}_{\rm F})|^2}$ and
$D_+(\varepsilon _n,\omega_m) \equiv D(\varepsilon _n + \omega _m) + D(\varepsilon _n)$.

For time-reversal invariant ground states with ${\bm d}\in \mathbb{R}^3$, the diagonal component of the quasiclassical Keldysh propagator becomes
\begin{align}
\delta g^-_0 = &
\frac{\omega \eta}{\omega^2-\eta^2}(1-\lambda)\Sigma_0^+\nn \\
&+\left\{1 + \frac{\eta^2}{\omega^2-\eta^2}(1-\lambda)\right\}\delta \Sigma _0^-
+\frac{1}{2} \eta \bar{\lambda}
d_{\mu}\delta d_{\mu}^- ,
\label{eq:deltag0}
\end{align}
and 
\begin{align}
\delta {\bm g}^+ =& \frac{\omega^2}{\omega^2-\eta^2}(1-\lambda){\bm \Sigma}^+
+ \bar{\lambda}({\bm d}\cdot {\bm \Sigma}^+) {\bm d} \nn \\
& +\frac{\omega \eta}{\omega^2-\eta^2}(1-\lambda){\bm \Sigma}^-
- \frac{i}{2}\omega\bar{\lambda}{\bm d}\times \delta{\bm d}^+,
\label{eq:deltag+}
\end{align}
where $\delta d^{\pm}_{\mu}\equiv \delta d_{\mu}\pm \delta d_{\mu}^{\ast}$, $\delta g^{\pm}_0\equiv\delta g_0\pm \delta \bar{g}_0$, and $\delta {\bm g}^{\pm}\equiv\delta {\bm g}\pm \delta \bar{\bm g}$. Similarly, the anomalous Keldysh propagators, $\delta f^{\pm}_{\mu}=\delta f_{\mu}\pm \delta \bar{f}_{\mu}$, are given by 
\begin{align}
\delta {\bm f}^- =& \left\{\frac{\gamma}{2} + \frac{1}{4}\bar{\lambda}(\omega^2-4|{\bm d}({\bm p}_{\rm F})|^2-\eta^2) \right\} 
\delta {\bm d}^-
+ \bar{\lambda}{\bm d} [ {\bm d}\cdot\delta {\bm d}^- ] \nn \\
& - \frac{1}{2}\bar{\lambda}{\bm d} \left( \eta  \delta \Sigma^-_0
+\omega \delta \Sigma^+_0\right),
\label{eq:f-}
\end{align}
and 
\begin{align}
\delta {\bm f}^+ =& \left\{\frac{\gamma}{2} + \frac{1}{4}\bar{\lambda}(\omega^2-\eta^2) \right\} 
\delta {\bm d}^- 
- \bar{\lambda}{\bm d} [ {\bm d}\cdot\delta {\bm d}^+ ] \nn \\
& + \frac{i}{2}\bar{\lambda}\left\{ \eta ({\bm d}\times \delta {\bm \Sigma}^-)
+ \omega ({\bm d}\times \delta {\bm \Sigma}^+) \right\}.
\label{eq:f+}
\end{align}
Here we introduce the bosonic response functions
\begin{gather}
\gamma (\omega _m) = \pi T \sum _{|\varepsilon _n|<\varepsilon _{n,{\rm c}}}\left[
\frac{1}{D(\varepsilon _n + \omega _m)} + 
\frac{1}{D(\varepsilon _n)}
\right], \\
\lambda(\omega _m) =  \pi T \sum _{|\varepsilon _n|<\varepsilon _{n,{\rm c}}} \frac{1}{D_+D(\varepsilon _n + \omega _m)D(\varepsilon _n)}.
\end{gather}
Analytic continuation to real frequencies of $\gamma(\omega_m)$ in the manner of 
Eq.~\eqref{eq:analytic} leads to 
\beq
\gamma({\bm p}_{\rm F}) = 2\int^{\varepsilon _{\rm c}}_{|{\bm d}|}d\varepsilon \frac{1}{\sqrt{\varepsilon^2-|{\bm d}|^2}} \tanh\left(\frac{\varepsilon}{2T}\right),
\eeq
where the frequency dependence can be ignored for $\omega_m \ll \varepsilon_{\rm c}$. The $\gamma$-function is related to the equilibrium gap equation \eqref{eq:gap0}
\beq
\frac{1}{V^{(E_u)}} = \frac{1}{2}\left\langle \gamma({\bm p}_{\rm F})|{\bm d}^{(E_{u})}_1({\bm p}_{\rm F})|^2\right\rangle _{\rm FS}.
\label{eq:Vgamma}
\eeq

The analytic continuation of $\lambda(\omega_m)$ yields the generalized Tsuneto function, $\lambda({\bm p}_{\rm F},{\bm Q},\omega)=|{\bm d}({\bm p}_{\rm F})|^2 \bar{\lambda}({\bm p}_{\rm F},{\bm Q},\omega)$, 
\begin{align}
{\lambda}
=& |{\bm d}|^2\int^{\varepsilon _{\rm c}}_{|{\bm d}|} d\varepsilon 
\frac{2\tanh(\varepsilon/2T)}{\sqrt{\varepsilon^2-|{\bm d}|^2}} 
\bigg[
\frac{\eta^2-2\omega\varepsilon _+}{(4\varepsilon^2_+-\eta^2)(\omega^2-\eta^2)+4\eta^2|{\bm d}|^2} \nn \\
&
+
\frac{\eta^2+2\omega\varepsilon _-}{(4\varepsilon^2_--\eta^2)(\omega^2-\eta^2)+4\eta^2|{\bm d}|^2}
\bigg],
\end{align}
where $\varepsilon _{\pm}\equiv \varepsilon \pm \omega/2$. The generalized Tsuneto function represents the ``stiffness'' of the condensate at the temperature $T$ and frequency $\omega$ and yields the momentum dependence in the nematic state. In the long-wavelength limit, the Tsuneto function reduces to
\begin{align}
\lambda({\bm p}_{\rm F},\omega) = \int^{\varepsilon _{\rm c}}_{|{\bm d}({\bm p}_{\rm F})|}
\frac{|{\bm d}({\bm p}_{\rm F})|^2d\varepsilon}{\sqrt{\varepsilon^2-|{\bm d}({\bm p}_{\rm F})|^2}}
\left\{
\frac{\tanh(\varepsilon/2T)}{\varepsilon^2-\omega^2/4}
\right\}+O(\eta^2),
\end{align}
with $\varepsilon _{\rm c}\rightarrow \infty$.
By introducing new variable, $\xi=\sqrt{\varepsilon^2-|{\bm d}|^2}$ and utilizing the series expansion 
\beq
\frac{\tanh(x/2T)}{2x} = T\sum _{\varepsilon _n} \frac{1}{\varepsilon^2_n+x^2},
\eeq
the Tsuneto function is rewritten in terms of the Matsubara sum 
\begin{align}
\lambda({\bm p}_{\rm F},\omega)=& \frac{\pi}{2}\left( 
\frac{2|{\bm d}({\bm p}_{\rm F})|}{\omega}
\right)\frac{\tanh(\omega/4T)}{\sqrt{1-(\omega/2|{\bm d}({\bm p}_{\rm F})|^2)}} \nn \\
&-T\sum _{|\varepsilon_n|<\varepsilon_{n,{\rm c}}} \frac{\pi|{\bm d}({\bm p}_{\rm F})|^2}{(\varepsilon^2_n+\omega^2/4)\sqrt{\varepsilon^2_n+|{\bm d}({\bm p}_{\rm F})|^2}},
\end{align}
for $\omega < 2|{\bm d}({\bm p}_{\rm F})|$. Above the pair-breaking frequency, $\omega > 2|{\bm d}({\bm p}_{\rm F})|$, the Tsuneto function acquires an imaginary part
\begin{align}
\lambda({\bm p}_{\rm F},\omega)=&
-T\sum _{|\varepsilon_n|<\varepsilon_{n,{\rm c}}} \frac{\pi|{\bm d}({\bm p}_{\rm F})|^2}{(\varepsilon^2_n+\omega^2/4)\sqrt{\varepsilon^2_n+|{\bm d}({\bm p}_{\rm F})|^2}}\nn \\
&+i\frac{\pi}{2}\left( 
\frac{2|{\bm d}({\bm p}_{\rm F})|}{\omega}
\right)\frac{\tanh(\omega/4T)}{\sqrt{(\omega/2|{\bm d}({\bm p}_{\rm F})|^2)-1}}.
\end{align}
The imaginary part, ${\rm Im}\lambda > 0$, reflects the density of pair excitations of Bogoliubov quasiparticles and gives rise to the damping of collective modes.

In the zero-temperature, long-wavelength limit, the Tsuneto function is recast into
\beq
\lambda(x) = \frac{\sin^{-1}(x)}{x\sqrt{1-x^2}},
\eeq
for $|x|\equiv |\omega/2|{\bm d}({\bm p}_{\rm F})||<1$ and 
\beq
\lambda(x) =
-\frac{\ln(x+\sqrt{x^2-1})}{x\sqrt{x^2-1}}+\frac{i\pi}{2x\sqrt{x^2-1}}
\eeq
for $|x|>1$.

\section{S2. Order parameter fluctuations}

For time-reversal invariant superconductors with ${\bm d}\in \mathbb{R}^3$, the order parameter fluctuations, $\delta d_{\mu}\equiv \delta d_{\mu}({\bm p}_{\rm F},{\bm Q},\omega)$, are obtained from the non-equilibrium gap equations
\beq
\hspace*{-7mm}
\delta d_{\mu}({\bm p}_{\rm F},{\bm Q},\omega) = -\bigg\langle \int\frac{d\varepsilon}{4\pi i}
V_{\mu \nu}({\bm p}_{\rm F},{\bm p}_{\rm F}^{\prime}) \delta f_{\nu}({\bm p}_{\rm F}^{\prime},\varepsilon;{\bm Q},\omega)
\bigg\rangle _{\rm FS},
\eeq
and Eqs.~\eqref{eq:f-} and \eqref{eq:f+} as
\begin{align}
\delta d^-_{\mu} =& - \bigg\langle
V_{\mu\nu}({\bm p}_{\rm F},{\bm p}_{\rm F}^{\prime})
\bigg\{
\frac{\gamma}{2}\delta d^-_{\nu}
+ \frac{1}{4}\bar{\lambda}
\left(
\omega^2-4|{\bm d}|^2 - \eta^2
\right)\delta d_{\nu}^- \nn \\
& + \bar{\lambda}d_{\nu}d_{\eta}\delta d^-_{\eta}
-\frac{1}{2}\bar{\lambda}d_{\nu}\left( \eta \delta \Sigma_0^-
+ \omega \delta \Sigma_0^+\right)
\bigg\}
\bigg\rangle^{\prime}_{\rm FS},
\label{eq:d}
\end{align}
and 
\begin{align}
\delta d^+_{\mu} =& - \bigg\langle
V_{\mu\nu}({\bm p}_{\rm F},{\bm p}_{\rm F}^{\prime})
\bigg\{
\frac{\gamma}{2}\delta d^+_{\nu}
+ \frac{1}{4}\bar{\lambda}
\left(
\omega^2 - \eta^2
\right)\delta d_{\nu}^+ 
- \bar{\lambda}d_{\nu}d_{\eta}\delta d^+_{\eta} \bigg\}\nn \\ 
& +\frac{i}{2}\bar{\lambda} \epsilon _{\nu\eta\tau} \left( 
\eta d_{\nu}\delta\Sigma_{\tau}^-
+ \omega d_{\nu}\delta\Sigma_{\tau}^+
\right)\bigg\rangle^{\prime}_{\rm FS},
\label{eq:d+}
\end{align}
The last terms in Eq.~\eqref{eq:d} and \eqref{eq:d+} represents an external source field that drives the order parameter fluctuations. 

Consider the $E_u$ state represented in Eq.~\eqref{eq:dEu}. We now expand the order parameter fluctuation in terms of the $(\Gamma,j)$ basis functions, ${\bm d}^{(\Gamma)}_j({\bm p}_{\rm F})$, as 
\beq
\delta d_{\nu} ({\bm p}_{\rm F},{\bm Q},t) = \sum^{\rm odd}_{\Gamma} \sum^{n_{\Gamma}}_{j=1} \mathcal{D}^{\rm C}_{\Gamma,j}({\bm Q},t) {\bm d}^{(\Gamma)}_j({\bm p}_{\rm F})
\eeq
where $\Gamma = \{A_{1u},A_{2u},E_u\}$ are the odd-parity irreducible representation of the crystal symmetry $D_{3d}$ and $n_{\Gamma}$ is the dimension of $\Gamma$. Substituting this into Eqs.~\eqref{eq:d} and \eqref{eq:d+} and ignoring the external source terms, one obtains the equations of motion for ${\rm C}={\pm}$ normal modes
\beq
\left[ \frac{\bar{\lambda}_{\Gamma,j}}{4}\omega^2 - \mathbb{M}^{\rm C}_{\Gamma,j}({\bm Q},\omega)\right] \mathcal{D}^{\rm C}_{\Gamma,j} ({\bm Q},\omega)= 0,
\eeq
which are the nonlinear equations on $\omega$. The excitation gaps are determined by 
\begin{align}
\frac{\bar{\lambda}_{\Gamma,j}}{4}\omega^2 -\mathbb{M}^-_{\Gamma,j}\equiv
\bigg[&
-\frac{1}{V^{(\Gamma)}_j} + \left\langle 
\bigg\{
\frac{\gamma}{2} +
\frac{\bar{\lambda}}{4} (\omega^2-4|{\bm d}|^2-\eta^2)
\bigg\}|{\bm d}^{(\Gamma)}_j|^2
\right\rangle _{\rm FS}
\nn \\
& + \left\langle 
\bar{\lambda}|{\bm d}\cdot{\bm d}^{(\Gamma)}_j|^2
\right\rangle _{\rm FS} 
\bigg]=0,
\label{eq:eomd-}
\end{align}
for ${\rm C}=-$ and 
\begin{align}
\frac{\bar{\lambda}_{\Gamma,j}}{4}\omega^2 -\mathbb{M}^+_{\Gamma,j}({\bm Q},\omega)\equiv
\bigg[&
-\frac{1}{V^{(\Gamma)}_j} + \left\langle 
\bigg\{
\frac{\gamma}{2} +
\frac{\bar{\lambda}}{4} (\omega^2-\eta^2)
\bigg\}|{\bm d}^{(\Gamma)}_j|^2
\right\rangle _{\rm FS}
\nn \\
& - \left\langle 
\bar{\lambda}|{\bm d}\cdot{\bm d}^{(\Gamma)}_j|^2
\right\rangle _{\rm FS} 
\bigg]=0,
\label{eq:eomd+}
\end{align}
for ${\rm C}=+$, where we have introduced $\bar{\lambda}_{\Gamma,j} \equiv \langle \bar{\lambda}|{\bm d}^{(\Gamma)}_j|^2\rangle _{\rm FS}$.
Equations~\eqref{eq:eomd-} and \eqref{eq:eomd+} are recast into 
\begin{widetext}
\begin{align}
\frac{\bar{\lambda}_{\Gamma,j}}{4}\omega^2 - \left\{ \langle |{\bm d}_j^{(\Gamma)}|^2 \rangle _{\rm FS}\ln\frac{T}{T_{\rm c}}
-\pi T \sum _{n}\left\langle 
\frac{|{\bm d}^{({\Gamma})}_j|^2}{\sqrt{\varepsilon^2_n+|{\bm d}|^2}} - \frac{|{\bm d}^{({\Gamma})}_j|^2}{|\varepsilon _n|}
\right\rangle _{\rm FS}\right\}
-\langle \lambda|{\bm d}^{(\Gamma)}_j|^2\rangle _{\rm FS}
-\frac{1}{4}\langle \eta^2\bar{\lambda|}{\bm d}^{(\Gamma)}_j|^2\rangle _{\rm FS}
+ \langle 
\bar{\lambda}|{\bm d}\cdot{\bm d}^{(\Gamma)}_j|^2
\rangle _{\rm FS} 
+  x_{\Gamma,j}\langle |{\bm d}_j^{(\Gamma)}|^2 \rangle _{\rm FS} = 0,
\label{eq:eomd-2}
\end{align}
for ${\rm C}=-$ and 
\begin{align}
\frac{\bar{\lambda}_{\Gamma,j}}{4}\omega^2 - \left\{ \langle |{\bm d}_j^{(\Gamma)}|^2 \rangle _{\rm FS}\ln\frac{T}{T_{\rm c}}
-\pi T \sum _{n}\left\langle 
\frac{|{\bm d}^{({\Gamma})}_j|^2}{\sqrt{\varepsilon^2_n+|{\bm d}|^2}} - \frac{|{\bm d}^{({\Gamma})}_j|^2}{|\varepsilon _n|}
\right\rangle _{\rm FS}\right\}
-\frac{1}{4}\langle \eta^2\bar{\lambda|}{\bm d}^{(\Gamma)}_j|^2\rangle _{\rm FS}
- \langle 
\bar{\lambda}|{\bm d}\cdot{\bm d}^{(\Gamma)}_j|^2
\rangle _{\rm FS} 
+ x_{\Gamma,j} \langle |{\bm d}_j^{(\Gamma)}|^2 \rangle _{\rm FS} = 0,
\label{eq:eomd+2}
\end{align}
for ${\rm C}=+$.
\end{widetext}
Here we have introduced the parameter
\beq
x_{\Gamma,j} = \ln T^{(\Gamma,j)}_{\rm c}/T_{\rm c} < 0,
\eeq
which represents the splitting of the critical temperature of the $(\Gamma,j)$ attractive interaction channel relative to the that of the ground state $T_{\rm c}=T^{(E_u,1)}_{\rm c}$. 
The coupling constant, $1/V^{(E_u)}_j$, is obtained from Eq.~\eqref{eq:Vgamma} and related to the measurable quantity, $T_{\rm c}\equiv T^{E_u,1}_{\rm c}$ by solving Eq.~\eqref{eq:Vgamma} at $T=T_{\rm c}$. The coupling constants in $A_{1u}$ and $A_{2u}$ channels, $1/V^{(A_{1u})}_j$ and $1/V^{(A_{2u})}_j$, are related to $T^{(A_{1u})}_{\rm c}$ and $T^{(A_{2u})}_{\rm c}$ by the gap equations which are obtained from Eq.~\eqref{eq:Vgamma} by replacing $V^{(E_u)}$ with $V^{(A_{1u})}$ and $V^{(A_{2u})}$.
Although both the Fermi surface average of the $\gamma$ function and $1/V^{(\Gamma)}_j$ depend on the cutoff of the Matsubara frequencies, $\varepsilon _{{\rm c}}$, their ultraviolet divergences are canceled out by each other, when $\varepsilon _{{\rm c}}$ is sufficiently large. Therefore, the resulting equations \eqref{eq:eomd-2} and \eqref{eq:eomd+2} are free from divergence in $\varepsilon _{{\rm c}}$.

Equations \eqref{eq:eomd-2} and \eqref{eq:eomd+2} determine the eigenfrequencies of $(\Gamma,j)$ bosonic modes in the nematic ground state, where we consider ${\bm d}({\bm p}_{\rm F})=\Delta {\bm d}^{(E_u)}_1({\bm p}_{\rm F})$ with $(\eta_1,\eta_2)=(1,0)$ without loss of generality. We note that for $(\Gamma,j)=(E_u,1)$, Eq.~\eqref{eq:eomd-2} has a pole at $Q=0$, i.e., ${\bm Q}={\bm 0}$ and $\omega = 0$
\beq
\omega^2 = \frac{\langle (\hat{\bm v}_{\rm F}\cdot\hat{\bm Q})^2\bar{\lambda}({\bm p}_{\rm F})|{\bm d}^{(E_u)}_1|^2\rangle_{\rm FS}}
{\langle \bar{\lambda}({\bm p}_{\rm F})|{\bm d}^{(E_u)}_1|^2\rangle _{\rm FS}} v^2_{\rm F}Q^2 ,
\eeq
corresponding to the NG mode associated with ${\tt U(1)_N}$ symmetry breaking. Here we have introduced $\hat{\bm v}_{\rm F}\equiv {\bm v}_{\rm F}/|{\bm v}_{\rm F}|$ and $\hat{\bm Q}\equiv{\bm Q}/|{\bm Q}|$. 

\begin{figure}[b!]
\includegraphics[width=85mm]{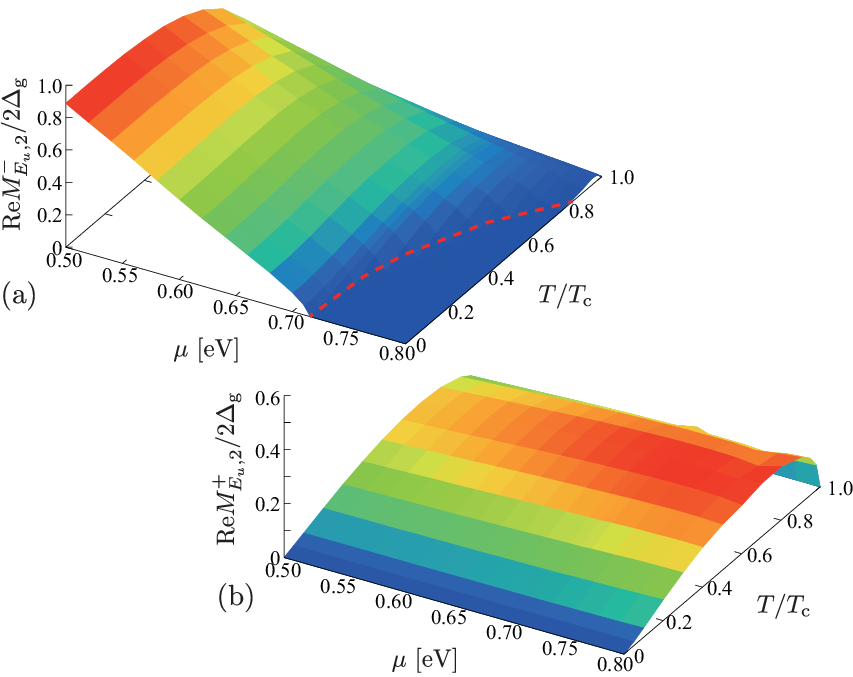}
\caption{Mass gap of the chilarity mode ${\rm Re}M^{-}_{E_u,2}$ (a) and the nematicity mode ${\rm Re}M^{+}_{E_u,2}$ (b) for $x_{E_u,2}=0$. The dashed curve corresponds to the dynamical instability of the chirality mode at which the mass gap closes, ${\rm Re}M^{-}_{E_u,2}=0$.}
\label{fig:s2}
\end{figure}

\begin{figure}[b!]
\includegraphics[width=85mm]{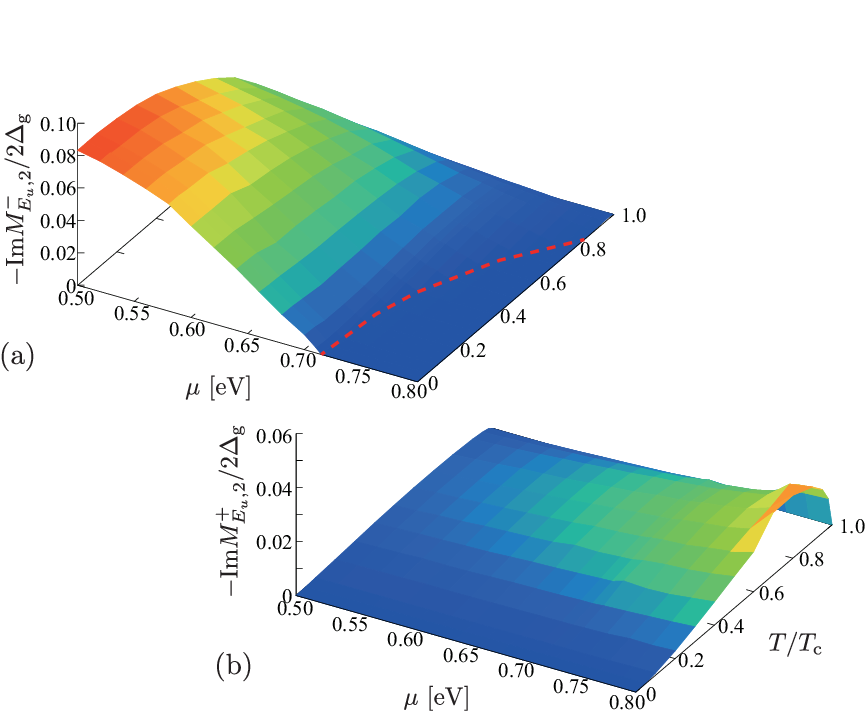}
\caption{Damping rate of the chilarity mode $-{\rm Im}M^{-}_{E_u,2}$ (a) and the nematicity mode $-{\rm Im}M^{+}_{E_u,2}$ (b) for $x_{E_u,2}=0$. The dashed curve corresponds to the dynamical instability of the chirality mode at which the mass gap closes, ${\rm Re}M^{-}_{E_u,2}=0$.}
\label{fig:s3}
\end{figure}

Figure~\ref{fig:s2} shows the temperature and chemical potential dependences of the mass gaps of the chilarity mode ${\rm Re}M^{-}_{E_u,2}$ and the nematicity mode ${\rm Re}M^{+}_{E_u,2}$. Here we set $x_{E_u,2}=0$, {\it i.e.}, $T^{E_u,2}_{\rm c}=T^{E_u,1}_{\rm c}=T_{\rm c}$. The nematicity mode acquires the finite mass gap at finite temperatures. The $T$-dependence resembles that of the ``normal-flapping mode'' in the superfluid $^3$He-A which is a vibration of the nodal direction about its equilibrium. The effective mass of such vibration mode is attributed to viscosity associated with the quasiparticle distribution around the gap nodes. Reflecting the density of the quasiparticles, the mass gap of the nematicity mode goes to zero with decreasing $T$. The damping rates of both the chirality and nematicity modes are shown in Fig.~\ref{fig:s3}. We find that the damping rates satisfy $-{\rm Im}M/{\rm Re}M<0.2$ for $T<T_{\rm c}$ for all $\mu$ and thus the chirality (nematicity) mode involves the stable vibration of the chirality of the Cooper pairs (the nodal direction or the nematicity angle) as long as ${\rm Re}M>0$. The increase of the damping rate of the nematicity mode with increasing $\mu$ reflects the increase of the quasiparticle density due to the evolution of the gap structure from the point nodes to line nodes.

\section{S3. Current response and power absorption}

We now consider the response to an electromagnetic field, 
\beq
\delta \Sigma _0({\bm p}_{\rm F},{\bm Q},\omega) = - \frac{e}{c}{\bm v}_{\rm F}\cdot{\bm A}({\bm Q},\omega), 
\eeq
where ${\bm A}({\bm Q},\omega)$ is the vector potential. %
For the electromagnetic response of unconventional (spin-triplet) superconductors we focus on the collisionless regime, where impurity vertex corrections may be neglected. The current response contains contributions from the bosonic collective modes driven by the electromagnetic field, in addition to Bogoliubov quasiparticle contributions to the current. The current response is obtained from the diagonal component of the quasiclassical Keldysh propagator as
\beq
{j}_{\mu}(Q) = -2e N_{\rm F} \left\langle {v}^{\mu}_{\rm F}\int \frac{d\epsilon}{4\pi i}
\delta{g}_0^-(\hat{\bm p}_{\rm F},{\bm Q};\varepsilon,\omega)
\right\rangle_{\rm FS}.
\eeq
The factor $2$ originates in the spin degeneracy. Substituting the solution in Eq.~\eqref{eq:deltag0}, one reads
\begin{align}
 {j}_{\mu}(Q) 
&\equiv 
K^{\rm QP}_{\mu\nu}(Q)A_{\nu} + K^{\rm CM}_{\mu\nu}(Q)
 A_{\nu}  .
\label{eq:current}
\end{align}
The first term in Eq.~\eqref{eq:current} is the contribution of quasiparticle (single-particle) excitations to the current 
\begin{align}
K^{\rm QP}_{\mu\nu}(Q)=-\frac{2e^2N_{\rm F}}{c}\left\langle\left\{
1 + \frac{\eta^2}{\omega^2-\eta^2}(1-\lambda)
\right\}{v}^{\mu}_{{\rm F}}{v}^{\nu}_{{\rm F}}\right\rangle _{\rm FS}.
\end{align}
The another term in the current response is attributed to the contribution of the bosonic collective excitations.
For time-reversal invariant superconducting states within the quasiclassical approximation ($T_{\rm c}\ll T_{\rm F}$), the electromagnetic wave can couple only to the ${\rm C}=-$ modes, including the chirality mode in the nematic state. The current response is decomposed into contributions from chiral $E_u$, $A_{1u}$, and $A_{2u}$ modes
\begin{align}
K^{\rm CM}_{\mu\nu}(Q)=K^{(E_u,1)}_{\mu\nu}(Q)+K^{(E_u,2)}_{\mu\nu}(Q)+K^{(A_{1u})}_{\mu\nu}(Q)+K^{(A_{2u})}_{\mu\nu}(Q),
\end{align}
where $K^{(\Gamma,j)}_{\mu\nu}(Q)$ represents contributions from $\mathcal{D}^-_{\Gamma,j}$ modes in the $(E_u,1)$ nematic ground state to the response function,
\begin{align}
K^{(\Gamma,j)}_{\mu\nu}({\bm Q},\omega) = -2e N_{\rm F} Q_{\tau}\sum^{n_{\Gamma}}_{j=1}\zeta^{(\Gamma,j)}_{\mu\tau} ({\bm Q},\omega)\frac{\delta\mathcal{D}^-_{\Gamma,j}({\bm Q},\omega)}{\delta A_{\nu}}.
\end{align}
We now introduce the symmetric tensor
\beq
\zeta^{(\Gamma,j)}_{\mu\nu}(Q)=\Delta
\left\langle \bar{\lambda}({\bm p}_{\rm F},{\bm Q},\omega)v^{\mu}_{\rm F} v^{\nu}_{\rm F} 
{\bm d}^{(E_u)}_{1}({\bm p}_{\rm F})\cdot{\bm d}^{(\Gamma)}_{j}({\bm p}_{\rm F})
\right\rangle _{\rm FS},
\eeq
which determines the coupling of transverse EM waves with ${\bm Q}$ and ${\bm A}$ to the bosonic modes, $\mathcal{D}^-_{\Gamma^{\prime}}$, in the ${\bm d}^{\Gamma}_i$ ground state. 
By using this tensor, the equation of motion for $\mathcal{D}^{-}_{\Gamma,i}$ is given as
\beq
\frac{\delta \mathcal{D}^{-}_{\Gamma,j}}{\delta A_{\mu}} = \left(\frac{e}{c}\right)\frac{Q_{\nu}\zeta _{\mu\nu}^{(\Gamma,i)}(Q)}{(\bar{\lambda}_{\Gamma,j}/4)\omega^2-\mathbb{M}^-_{\Gamma,j}(Q)}.
\eeq
The zeros of the denominator correspond to Eq.~\eqref{eq:eomd-} and determine the eigenfrequencies of $(\Gamma,j)$ bosonic modes in the nematic ground state (see also Eq.~(8) in the main text). To this end, the response function, $K^{\rm CM}_{\mu\nu}(Q)$, reduces to 
\beq
K^{\rm CM}_{\mu\nu}(Q) = - \frac{e^2}{c}N_{\rm F}\sum^{\rm odd}_{\Gamma}\sum^{n_{\Gamma}}_{j=1}
\frac{Q_{\tau}Q_{\eta}\zeta^{(\Gamma,j)}_{\mu\tau}(Q)
\zeta^{(\Gamma,j)}_{\nu\eta}(Q)}{(\bar{\lambda}_{\Gamma,j}/4)\omega^2-\mathbb{M}^-_{\Gamma,j}(Q)}.
\eeq

As we mention in the main text, $\zeta$ is subject to the symmetries of the equilibrium order parameter (${d}^{E_u}_1$) and dynamical bosonic field ($d^{\Gamma}_i$). For the $E_{u,1}$ ground state, nontrivial components are $\zeta _{xy}=\zeta _{yx}$ for the chiral $E_u$ mode ($(\Gamma,i)=(E_u,2)$) and $\zeta _{xz}=\zeta _{zx}$ for the chiral $A_{2u}$ mode ($\Gamma=As_{2u}$). As $\mathbb{M}(Q)$ is subject to the enlarged ${\tt D_{\infty}}$ symmetry, the contribution from the chiral $E_u$ mode to the response function is
\beq
K^{(E_u,2)}_{\mu\nu}(Q) = \tilde{K}^{(E_u,2)}(Q)\begin{pmatrix}
\hat{Q}^2_y & \hat{Q}_x\hat{Q}_y & 0 \\
\hat{Q}_x\hat{Q}_y & \hat{Q}^2_x & 0 \\
0 & 0 & 0
\end{pmatrix}.
\eeq
Similarly, the contribution from the chiral $A_{2u}$ mode reduces to 
\beq
K^{(A_{2u})}_{\mu\nu}(Q) = \begin{pmatrix}
\tilde{K}^{(A_{2u})}_{xx}(Q) \hat{Q}^2_z & 0 & \tilde{K}^{(A_{2u})}_{xz}(Q) \hat{Q}_x\hat{Q}_z\\
0 & 0 & 0 \\
\tilde{K}^{(A_{2u})}_{zx}(Q) \hat{Q}_z\hat{Q}_x & 0& \tilde{K}^{(A_{2u})}_{zz}(Q) \hat{Q}^2_x \\
\end{pmatrix}.
\eeq
%
The coupling of transverse EM fields to chiral $A_{1u}$ modes is accidentally prohibited by the enlarged ${\tt D_{\infty}}$ symmetry, i.e., $K^{(A_{1u})}_{\mu\nu}=0$. 

The signatures of the collective mode spectrum are captured by the the EM power absorption, which we calculate following the scheme in Ref.~\onlinecite{hirschfeld}. Consider a metal-vacuum interface at $z=0$, where $\hat{\bm z}\parallel{\bm Q}$ denotes the direction of the electromagnetic wave propagation is normal to the interface. The power absorption is obtained from Joule's law by integrating the energy density dissipated over the half space of the metal.
\beq
P(\omega) = \int^{\infty}_0 dz\, {\rm Re}\left[{\bm E}^{\ast}({\bm r},\omega)\cdot {\bm j}({\bm r},\omega)\right].
\eeq

\begin{figure}[t!]
\includegraphics[width=85mm]{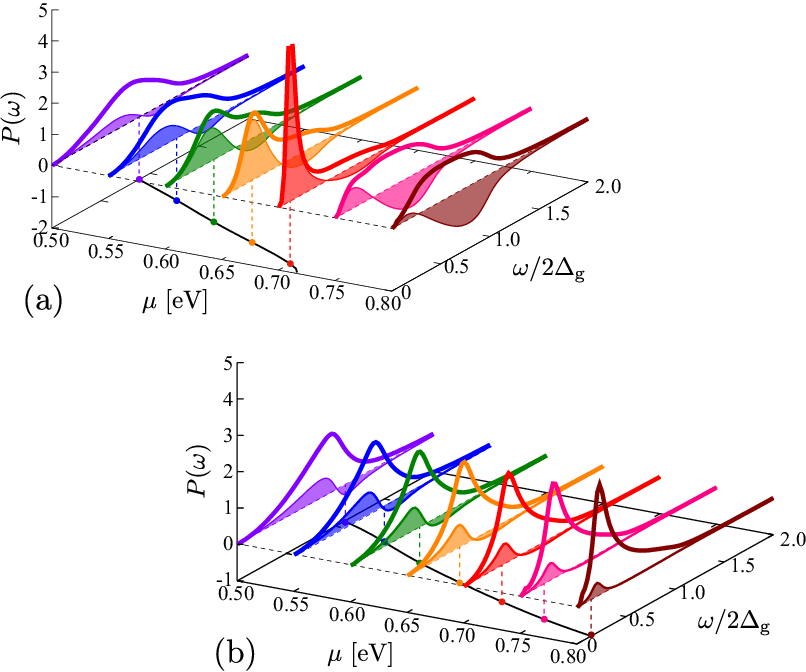}
\caption{Power absorption spectra, $P(\omega)$, in the nematic state with ${\bm d}({\bm p}_{\rm F})=\Delta(T,\mu){\bm d}^{(E_u)}_1({\bm p}_{\rm F})$ for $T=0.05T_{\rm c}$: (a) ${\bm A}\parallel {\bm x}$ and ${\bm Q}\parallel{\bm y}$ and (b) ${\bm A}\parallel {\bm z}$ and ${\bm Q}\parallel{\bm y}$. The shaded are stands for the contributions of bosonic excitations, $P^{\rm CM}(\omega)$. The mass gap of the chirality mode, $M^-_{E_u,2}$, and the $A_{2u}$ mode, $M^-_{A_{2u}}$, are also shown in (a) and (b), respectively. We set $x_{Eu,2}=0$ and $x_{A_{1u}}=x_{A_{2u}}=-1.5$, corresponding to $T^{(E_u,1)}_{\rm c}=T^{(E_u,2)}_{\rm c}=T_{\rm c}$ and $T^{(A_{1u})}_{\rm c}=T^{(A_{2u})}_{\rm c}= 0.22T_{\rm c}$.}
\label{fig:power2}
\end{figure}

Following Refs.~\onlinecite{hirschfeld} and \onlinecite{yip}, we map the half-space boundary-value problem with ${\bm B}(z=0)={\bm B}_0$ onto a full-space Maxwell equation with specular boundary condition at the interface. The electrons passing through the interface experience the mirro-reflected vector potential and magnetic field, as ${\bm A}(z)={\bm A}(-z)$ and ${\bm B}(z)=-{\bm B}(-z)$. Hence, the full-space Maxwell equation is accompanied by an external current sheet associated with the discontinuity of the field at the interface, ${\bm j}^{\rm ext}(\omega) = -\frac{c}{2\pi}B_0(\omega)$,
\beq
\left( 
\delta _{\mu\nu} 
{\bm \nabla}^2 
+ \frac{4\pi}{c}K_{\mu \nu} \right) A_{\nu}
= -\frac{4\pi}{c}{ j}^{\rm ext}_{\mu},
\eeq
where $K_{\mu\nu}\equiv K^{\rm QP}_{\mu\nu}+K^{\rm CM}_{\mu\nu}$ is the response function defined in Eq.~\eqref{eq:current}. 

Let us consider the response of the $E_{u,1}$ ground state to transverse EM fields with ${\bm Q}\parallel \hat{\bm x}$. The current response is given as $\delta{\bm j}/\delta A=(K^{\rm QP}+\tilde{K}^{E_{u,2}})\hat{\bm y}\equiv K\hat{\bm y}$ for ${\bm A}\parallel\hat{\bm y}$ and $\delta{\bm j}/\delta A=(K^{\rm QP}+\tilde{K}^{A_{2u}})\hat{\bm z}\equiv K\hat{\bm z}$ for ${\bm A}\parallel\hat{\bm z}$.
Solving for the Fourier component ${\bm A}({\bm Q},\omega)$, we have 
\beq
A({\bm Q},\omega) = - \frac{2B_0(\omega)}{{\bm Q}^2-(\frac{4\pi}{c})K({\bm Q},\omega)}.
\eeq
Thus, the power absorption is given in terms of the response function as
\begin{align}
P(\omega) = -\frac{2\omega |B_0(\omega)|^2}{c} \int^{\infty}_0 \frac{dQ}{2\pi} 
\frac{{\rm Im}K(Q,\omega)}{|Q^2-(\frac{4\pi}{c})K(Q,\omega)|^2}.
\label{eq:P}
\end{align}
%
Here we introduce the London penetration depth at $T=0$
\beq
\Lambda = \sqrt{\frac{mc^2}{4\pi ne^2}}.
\eeq
Then, the power absorption in Eq.~\eqref{eq:P} is 
\beq
P(\omega) = -\frac{\omega |B_0(\omega)|^2}{2\pi}\Lambda^2 \int^{\infty}_0 \frac{dQ}{2\pi} 
\frac{{\rm Im}K(Q,\omega)}{|(Q\Lambda)^2-\tilde{K}(Q,\omega)|^2}.
\eeq
The kernel is $\tilde{K}\approx {\rm Re}(\tilde{K}) \approx -1$. The power absorption then becomes
\beq
P(\omega) \approx -\frac{\omega |B_0(\omega)|^2}{2\pi}\Lambda^2 \int^{\infty}_0 \frac{dQ}{2\pi} 
\frac{{\rm Im}K(Q,\omega)}{|(Q\Lambda)^2+1|^2}.
\eeq
Hence the power absorption reduces to the average of the dissipation part of the current kernel over the penetration depth $Q \lesssim \Lambda^{-1} < \xi^{-1}$.
The quasiparticle and collective mode contributions of the power absorption are given as 
\begin{gather}
P^{\rm QP}(\omega) \approx -\frac{\omega |B_0(\omega)|^2}{2\pi}\Lambda^2 \int^{\infty}_0 \frac{dQ}{2\pi} 
\frac{{\rm Im}K^{\rm QP}(Q,\omega)}{|(Q\Lambda)^2+1|^2}, \\
P^{\rm CM}(\omega) \approx -\frac{\omega |B_0(\omega)|^2}{2\pi}\Lambda^2 \int^{\infty}_0 \frac{dQ}{2\pi} 
\frac{{\rm Im}K^{\rm CM}(Q,\omega)}{|(Q\Lambda)^2+1|^2}.
\end{gather}

In Fig.~\ref{fig:power2}, we plot power absorption spectra, $P(\omega)$, in the nematic state at $T=0.05T_{\rm c}$. Figure~\ref{fig:power2} (a) and \ref{fig:power2} (b) show the absorption of the transverse EM wave with $({\bm A}\parallel {\bm x}, {\bm Q}\parallel{\bm y})$ and $({\bm A}\parallel {\bm z}$ and ${\bm Q}\parallel{\bm y})$, respectively. The former case resonates the chirality mode $\mathcal{D}^-_{E_u,2}$, while the latter involves the resonance of the massive $A_{2u}$ mode, $\mathcal{D}^-_{A_{2u}}$, where the shaded are represents the contributions of bosonic excitations, $P^{\rm CM}(\omega)$.

\section{S4. Dynamical Spin Susceptibilities}

In the quasiclassical theory which is reliable in the weak coupling limit $T_{\rm c}/T_{\rm F}\ll 1$, the ${\rm C}=+$ sector of the bosonic excitations including the nematicity mode cannot be coupled to transverse EM waves. We here demonstrate the impact of the nematicity vibration mode on the dynamical magnetic response of the nematic state. Let us consider a time-dependent field directly coupled to the magnetic moment of the electrons in nematic superconductors (e.g., rf-fields). The quasiclassical self-energies for the dynamical Zeeman term is obtained by projecting Eq.~\eqref{eq:hz} onto the conduction band as 
\beq
\delta \Sigma_{\mu} ({\bm Q},\omega)=\frac{1}{1+F^{\rm a}_0}
\sum _{\mu=x,y,z} \frac{1}{2} {g}^{\rm eff}_{\mu} \mu_{\rm B} s_{\mu}{H}_{\mu}({\bm Q},\omega),
\eeq
where $F^{\rm a}_0$ is the Fermi liquid parameter associated with the anti-symmetric spin-dependent channel of the quasiparticle scattering process. The effective $g$-factor of the conduction band electrons is given by
\beq
g_i^{\rm eff} \equiv g_{i0} + g_{ix}\frac{m}{\sqrt{m^2+v^2_zf^2_z+v^2p_{\parallel}^2}}
\approx g_{i0} + g_{ix}.
\eeq
For the parent material, Bi$_2$Se$_3$, the effective $g$-factor is strongly anisotropic as $g^{\rm eff}_x=g^{\rm eff}_y = -8.24$ and $g^{\rm eff}_z = -50.8$.

The magnetic response of the system is described by the magnetization density which is obtained from the vectorial components of the diagonal propagators as
\begin{align}
M_{\mu}({\bm Q},\omega) =& {M}_{{\rm N},\mu}({\bm Q},\omega) \nn \\
&+\frac{g^{\rm eff}_{\mu}\mu_{\rm B}\mathcal{N}_{\rm F}}{1+F^{\rm a}_0}
\left\langle \int \frac{d\varepsilon}{4\pi i} \delta { g}_{\mu}({\bm p}_{\rm F},{\bm Q};\varepsilon,\omega)
\right\rangle _{\rm FS},
\label{eq:M}
\end{align}
where the first term in the right-hand side is the magnetization in the normal state, 
\beq
{ M}_{{\rm N},\mu}=\frac{1}{2}\frac{(g^{\rm eff}_{\mu}\mu_{\rm B})^2\mathcal{N}_{\rm F}}{1+F^{\rm a}_0}H_{\mu}.
\eeq 
Substituting Eq.~\eqref{eq:deltag+} into Eq.~\eqref{eq:M}, one finds that the dynamical spin susceptibilities are composed of two terms 
\beq
\chi _{\mu \nu} ({\bm Q},\omega)  \equiv \frac{\delta M_{\mu}}{\delta H_{\nu}} = 
\chi^{\rm QP}_{\mu\nu}({\bm Q},\omega)  + \chi^{\rm CM}_{\mu\nu}({\bm Q},\omega) .
\eeq
The first term corresponds to the contributions of Bogoliubov quasiparticles and equilibrium ${\bm d}$-vector,
\begin{align}
\chi^{\rm QP}_{\mu\nu} = \chi _{\rm N}\bigg[&
\delta _{\mu \nu}\left\{ 
1+\frac{1}{1+F^{\rm a}_0}\left\langle \frac{\omega^2}{\omega^2-\eta^2}(1-\lambda) \right\rangle _{\rm FS}
\right\} \nn \\
& + \frac{1}{1+F^{\rm a}_0}\left\langle \bar{\lambda} d_{\mu}d_{\nu}\right\rangle _{\rm FS}
\bigg].
\end{align}
This reduces to the spin susceptibility of the equilibrium nematic state at ${\bm Q}\rightarrow {\bm 0}$ and $\omega\rightarrow 0$.

For the nematic state of $M_x$Bi$_2$Se$_3$, only the diagonal components of the tensor remains nontrivial. The second term, $\chi^{\rm CM}_{\mu\nu}$, describes the resonance of the bosonic excitations in the ${\rm C}=+$ sector including the nematicity vibration mode, 
\beq
\chi ^{\rm CM}_{\mu\nu}= - \frac{i}{4}\omega \sum^{\rm odd}_{\Gamma}\sum^{n_{\Gamma}}_{j=1}
\left\langle
\bar{\lambda}\left( {\bm d}\times {\bm d}^{(\Gamma)}_j \right) _{\mu}
\right\rangle _{\rm FS}\left( \frac{\delta \mathcal{D}^+_{\Gamma,j}}{\delta H_{\nu}}\right).
\eeq
Similarly with the coupling of the chirality mode to transverse EM waves, the nematicity mode leads to a pronounced peak of dynamical spin susceptibilties at the resonant frequency,
\beq
\frac{\delta \mathcal{D}^+_{\Gamma,j}}{\delta H_{\nu}}
= -\frac{i}{2}g^{\rm eff}_{\nu}\mu_{\rm B}\omega\frac{\langle\bar{\lambda}({\bm d}\times{\bm d}^{(\Gamma)}_j)_{\nu}\rangle_{\rm FS}}{(\bar{\lambda}_{\Gamma,j}/4)\omega^2-\mathbb{M}^+_{\Gamma,j}}.
\eeq
The resonance of the nematicity mode to the magnetic response is subject to the selection rule due to $\langle\bar{\lambda}({\bm d}\times{\bm d}^{(\Gamma)}_j)\rangle_{\rm FS}$. For the nematicity mode, $\mathcal{D}^+_{E_u,2}$, one finds 
$\langle[{\bm d}^{(E_u)}_1\times {\bm d}^{(E_u)}_2]_{x}\rangle _{\rm FS} =\langle[{\bm d}^{(E_u)}_1\times {\bm d}^{(E_u)}_2]_{y}\rangle _{\rm FS}= 0$ and $\langle[{\bm d}^{(E_u)}_1\times {\bm d}^{(E_u)}_2]_{z}\rangle _{\rm FS} \neq 0$. This implies the selection rule that the nematicity mode contributes only to the longitudinal dynamical spin susceptibility 
\beq
\chi^{\rm CM}_{zz}\neq 0,
\eeq 
and otherwise $\chi^{\rm CM}_{\mu\nu}=0$. We will study in details the impact of the nematicity mode on dynamical spin susceptibilities elsewhere.


%
\end{document}